\documentclass[
prd,twocolumn,aps,superscriptaddress,nofootinbib]{revtex4-1}
\usepackage{amssymb}
\usepackage{amsmath}
\usepackage[colorlinks=true, pdfstartview=FitV, linkcolor=blue, citecolor=blue, urlcolor=blue]{hyperref}
\usepackage{bm}

\newcommand{\bx}{{\boldsymbol x}}
\newcommand{\bp}{{\boldsymbol p}}
\newcommand{\bq}{{\boldsymbol q}}
\newcommand{\bv}{{\boldsymbol v}}
\newcommand{\bu}{{\boldsymbol u}}
\newcommand{\bB}{{\boldsymbol B}}
\newcommand{\bJ}{{\boldsymbol J}}

\newcommand{\bOmega}{{\boldsymbol \Omega}}
\newcommand{\bomega}{{\boldsymbol \omega}}
\newcommand{\tbv}{\tilde{\boldsymbol v}}
\newcommand{\tepsilon}{\tilde{\epsilon}}
\newcommand{\tF}{\tilde{F}}

\newcommand{\hbp}{{\hat{\boldsymbol p}}}
\newcommand{\hbq}{{\hat{\boldsymbol q}}}
\newcommand{\hmu}{{\hat{\mu}}}
\newcommand{\hnu}{{\hat{\nu}}}
\newcommand{\hrho}{{\hat{\rho}}}
\newcommand{\hsigma}{{\hat{\sigma}}}

\newcommand{\halpha}{{\hat{\alpha}}}
\newcommand{\hbeta}{{\hat{\beta}}}

\newcommand{\hzero}{{\hat{0}}}
\newcommand{\hone}{{\hat{1}}}
\newcommand{\htwo}{{\hat{2}}}
\newcommand{\hthree}{{\hat{3}}}
\newcommand{\lD}{\overleftarrow{D}}
\newcommand{\lpart}{\overleftarrow{\partial}}
\newcommand{\lcalG}{\overleftarrow{\mathcal{G}}}

\newcommand{\calC}{\mathcal{C}}

\newcommand{\calG}{\mathcal{G}}
\newcommand{\calH}{\mathcal{H}}

\newcommand{\calF}{\mathcal{F}}
\newcommand{\calP}{\mathcal{P}}
\newcommand{\calV}{\mathcal{V}}
\newcommand{\calA}{\mathcal{A}}
\newcommand{\calS}{\mathcal{S}}
\newcommand{\calR}{\mathcal{R}}
\newcommand{\calL}{\mathcal{L}}
\newcommand{\calX}{\mathcal{X}}
\newcommand{\calY}{\mathcal{Y}}
\newcommand{\calZ}{\mathcal{Z}}
\newcommand{\calW}{\mathcal{W}}
\newcommand{\calUin}{\mathcal{U}_\text{in}}
\newcommand{\calUrot}{\mathcal{U}_\text{rot}}
\newcommand{\bbP}{\mathbb{P}}
\newcommand{\bbY}{\mathbb{Y}}
\newcommand{\bbT}{\mathbb{T}}
\newcommand{\bbM}{\mathbb{M}}

\newcommand{\intp}{\int_{\boldsymbol p}}
\newcommand{\feq}{f_\text{eq}}

\newcommand{\zero}{{(0)}}
\newcommand{\dis}{\displaystyle}

\renewcommand\b{\beta}
\renewcommand\d{\delta}

\renewcommand\l{\lambda}
\renewcommand\r{\rho}

\newcommand\e{\epsilon}

\newcommand\m{\mu}
\newcommand\n{\nu}

\newcommand\s{\sigma}

\newcommand\ola{\overleftarrow}

\begin{document}

\title{Chiral kinetic theory in curved spacetime}

\author{Yu-Chen~Liu}
\affiliation{Physics Department and Center for Particle Physics and Field Theory, Fudan University, \\ Shanghai 200433, China}
\author{Lan-Lan~Gao}
\affiliation{Physics Department and Center for Particle Physics and Field Theory, Fudan University, \\ Shanghai 200433, China}
\author{Kazuya~Mameda}
\email{mame@nt.phys.s.u-tokyo.ac.jp}
\affiliation{Physics Department and Center for Particle Physics and Field Theory, Fudan University, \\ Shanghai 200433, China}
\author{Xu-Guang~Huang}
\email{huangxuguang@fudan.edu.cn}
\affiliation{Physics Department and Center for Particle Physics and Field Theory, Fudan University, \\ Shanghai 200433, China}
\affiliation{Key Laboratory of Nuclear Physics and Ion-beam Application (MOE), Fudan University, \\ Shanghai 200433, China}

\begin{abstract}
Many-body systems with chiral fermions exhibit anomalous transport phenomena originated from quantum anomalies.
Based on quantum field theory, we derive the kinetic theory for chiral fermions interacting with an external electromagnetic field in a background curved geometry. The resultant framework is U(1) gauge invariant and local Lorentz and diffeomorphism covariant.
It is particularly useful to study the gravitational or non-inertial effects for chiral fermions. As the first application, we study the chiral dynamics in a rotating coordinate and clarify the roles of the Coriolis force and spin-vorticity coupling in generating the chiral vortical effect. We also show that the chiral vortical effect is an intrinsic phenomenon of a rotating chiral fluid, and thus independent of the observer's frame.
\end{abstract}

\maketitle

\section{Introduction}
Quantum anomaly is a prominent concept in the transport phenomena of chiral fermions. One of its most novel consequences is the generation of parity-breaking currents, typified by the chiral magnetic effect (CME)~\cite{Kharzeev:2007jp,Fukushima:2008xe} and chiral vortical effect (CVE)~\cite{Vilenkin:1979ui,Erdmenger:2008rm,Banerjee:2008th,Son:2009tf}. A crucial feature of these anomalous currents is that they are insensitive to the details of interactions and are thus universal. For this reason, such phenomena have received a lot of attention in a wide context of physics ranging from the high-energy nuclear physics~\cite{Kharzeev:2015znc,Huang:2015oca,Hattori:2016emy,Skokov:2016yrj} and astrophysics~\cite{Charbonneau:2009ax,Ohnishi:2014uea,Masada:2018swb} to condensed matter physics~\cite{Miransky:2015ava,Armitage:2017cjs,Burkov:2015hba}.

To study the real-time dynamics of the anomalous transport phenomena, the chiral kinetic theory (CKT) is a promising approach which is applicable when the system is dilute and the external fields are weak~\cite{Stephanov:2012ki,Son:2012zy,Gao:2012ix,Chen:2012ca,Huang:2015mga,Hidaka:2016yjf,Huang:2018wdl,Mueller:2017arw}. In CKT, the chiral anomaly is encoded through the Berry curvature~\cite{Berry:1984jv}, which modifies the Boltzmann equation and the phase space measure. Recently, various aspects of the CKT were investigated, including the Lorentz covariance~\cite{Chen:2014cla,Chen:2015gta,Hidaka:2016yjf,Huang:2018wdl}, consistent versus covariant anomalies~\cite{Gorbar:2016ygi,Carignano:2018gqt}, particle collisions~\cite{Chen:2015gta,Hidaka:2016yjf,Hidaka:2017auj,Hidaka:2018ekt}, etc.

Despite these developments, so far the CKT is restricted to flat spacetime and thus not conventional to explore the anomalous transport phenomena induced by gravitational or non-inertial effects~\cite{Erdmenger:2008rm,Banerjee:2008th,Landsteiner:2011cp,Golkar:2012kb,Golkar:2015oxw,Jensen:2012kj,Glorioso:2017lcn}. Although the classical Boltzmann equation is readily extended to curved spacetime, the formulation of the CKT in curved spacetime is highly nontrivial.
The very few attempts so far~\cite{Basar:2013qia,Dayi:2016foz,Huang:2018aly} considered only a special curved spacetime, that is, the rotating coordinate%
~\footnote{In this paper, the rotating coordinate will be regarded as a curved spacetime even though its Riemann curvature is zero, while the term ``flat spacetime'' is specifically referred to as the Minkowski spacetime.},
and lacked the diffeomorphism (i.e., the general coordinate transformation) covariance.
The more rigorous derivation should start from quantum field theory in curved spacetime.

In this paper, we derive the CKT in an arbitrary curved spacetime and external electromagnetic field, based on the Wigner function formalism that respects the U(1) gauge invariance, and the local Lorentz and diffeomorphism covariance~\cite{Winter:1986da,Calzetta:1987bw,Fonarev:1993ht}. We apply the resultant framework to a rotating coordinate and examine the frame dependence of the CVE, which is so far unclear. We show that, depending on the observer's frame, the Coriolis force and spin-vorticity coupling (and the side-jump effect) can be responsible to the generation of the CVE, but the total CVE current is always independent of the observer's frame.

Throughout this paper, we choose the unit $c=k_\text{B}=e=1$ (with $e$ the electric charge), but keep $\hbar$ explicit;
(un)hatted Greek indices denote local flat (curved) spacetime coordinates;
$\eta_{\halpha\hbeta} = \text{diag}(1,-1,-1,-1)$ is the Minkowski metric;
$\nabla_\mu$ denotes the covariant derivative with respect to the diffeomorphism and local Lorentz transformation, e.g., for scalars $\nabla_\mu f=\partial_\mu f$, for vectors $\nabla_\mu V_\nu = \partial_\mu V_\nu  - \Gamma^\lambda_{\mu\nu}V_\lambda$, for spinors $\nabla_\mu\psi=(\partial_\mu + \Gamma_\mu)\psi$, where the spin connection is $\Gamma_\mu= -\frac{i}{4}\sigma^{\hat{\alpha}\hat{\beta}} g_{\rho\sigma}e^\rho_\halpha(\partial_\mu e^\sigma_\hbeta + \Gamma^\sigma_{\mu\nu}e^\nu_{\hbeta})$, with the spin matrix $\sigma^{\halpha\hbeta} = \frac{i}{2}[\gamma^{\halpha},\gamma^{\hbeta}]$, vierbein $e^\mu_{\halpha}$, and the Christoffel symbol $\Gamma_{\mu\nu}^\rho = \Gamma_{\nu\mu}^\rho$;
the Levi-Civita symbol is $\varepsilon^{\m\n\r\s}=\sqrt{-g(x)}\varepsilon^{\hmu\hnu\hrho\hsigma}$ with $\varepsilon^{\hzero\hone\htwo\hthree} = -\varepsilon_{\hzero\hone\htwo\hthree} = 1$ and $g=\det (g_{\mu\nu})$;
the Dirac matrices satisfy $\{\gamma^\mu,\gamma^\nu\} = 2g^{\mu\nu}$ and $\gamma^5$ is defined as $\gamma^5= (-i/4!)\varepsilon_{\mu\nu\rho\sigma} \gamma^\mu\gamma^\nu\gamma^\rho\gamma^\sigma$.

\section{Phase space and horizontal lift}
In curved spacetime the definition of the phase space is subtle because a global notion of momentum is usually not permitted.
This means that for each position $x$ on the spacetime manifold $\bbM$, one can introduce a momentum space $\bbP_x$ attached to $\bbM$.
The phase space is then the collection of $(x,\bbP_x)$, which constitutes a fiber bundle.
One of the natural choices for $\bbP_x$ is the tangent or cotangent space so that the usual momentum space is reproduced in Minkowski spacetime.
In this paper, we employ the latter.
That is, we define the momentum variable $p_\mu$ as a point in the cotangent space (that is, $p_\mu$ is a covariant vector on $\bbM$), and the corresponding phase space is the cotangent bundle $\bbT^*\bbM$.
Similarly, a point $y^\mu$ in the tangent space $\bbY_x$ (that is, $y^\mu$ is a contravariant vector on $\bbM$) is defined as the position variable canonically conjugate to $p_\mu$.
The set of $(x,\bbY_x)$ builds the tangent bundle $\bbT\bbM$.

The Wigner function $W(x,p)$ for Dirac fermions that we will introduce in next section is required to transform covariantly under the $\text{U}(1)$ gauge transformation, local Lorentz transformation, and the diffeomorphism.
The $\text{U}(1)$ covariance is ensured when $W(x,p)$ is suitably constructed with $\partial_\mu + iA_\mu/\hbar$, instead of $\partial_\mu$, see next section and Ref.~\cite{Elze:1986qd,Heinz:1983nx}. Consequently, $W(x,p)$ is transformed as $W(x,p)\to S(x)W(x,p)S^{-1}(x)$, where $S$ is the representation matrix of  the $\text{U}(1)$ gauge transformation.
In a similar manner, the local Lorentz covariance can be kept by introducing the spin connection $\Gamma_\mu$ and replacing $\partial_\mu$ by the covariant derivative $\nabla_\mu$; $W(x,p)$ is thus transformed as a bispinor, $W(x,p)\to U(\Lambda)W(x,\Lambda p)U^{-1}(\Lambda)$, where $\Lambda$ is the local Lorentz transformation at $x$ and $U(\Lambda)$ is its spinorial representation.
The diffeomorphism covariance of $W(x,p)$ needs more careful treatment.
This is because the diffeomorphism affects functions in $\bbT\bbM$ or $\bbT^*\bbM$ in a very nontrivial way.
In fact, the proper covariant derivative on $\bbT^*\bbM$ is defined as follows~\cite{nakahara2003geometry,Fonarev:1993ht}:
\begin{equation}
\label{eq:lift}
 D_\mu = \nabla_\mu 
 + {\Gamma^\lambda_{\mu\nu}}p_\lambda\partial^\nu_p
\end{equation}
with $\partial^\mu_p=\partial/\partial p_\mu$.
The derivation of Eq.~\eqref{eq:lift} based on the parallel displacement is shown in Appendix \ref{app:horizontal}.
In differential geometry, such defined $D_\mu$ is called the horizontal lift of $\nabla_\m$ on $\mathbb M$ to $\bbT^*\bbM$.
Similarly to Eq.~\eqref{eq:lift}, the covariant derivative in $\bbT\bbM$ is defined by $D_\mu = \nabla_\mu - \Gamma^\lambda_{\mu\nu} y^\nu\partial^y_\lambda$.

The implementation of the horizontal lift brings a great advantage in analysis. That is, we can regard $p_\m$ and $y^\m$ as ``$x$-independent" variables under the parallel transport by $D_\m$, because of
\begin{equation}
 D_\mu p_\nu = D_\mu y^\nu = 0 \,.
\end{equation}
As a result, for an arbitrary function $\Psi(x)$ on $\mathbb M$, its lifted image in $\bbT\bbM$ is represented as the function translated by $D_\mu$:
$\Psi(x,y) \equiv \Psi(x)+y^\mu\nabla_\mu\Psi(x) +\frac{1}{2}y^\mu y^\nu\nabla_\mu\nabla_\nu\Psi(x)+\cdots = \exp(y\cdot D)\Psi(x)$.
Furthermore, the Fourier transformation from $\bbT\bbM$ to $\bbT^*\bbM$ is expressed as
$\Psi(x,p)= \int d^4y\,\sqrt{-g(x)}\, \exp(-ip\cdot y/\hbar)\,\Psi(x,y)$.

\section{Quantum transport in curved spacetime}
With the above preparation, we define the fermionic Wigner function covariantly under the U(1) gauge, local Lorentz transformations, and diffeomorphism, as follows:
\begin{equation}
 \label{eq:W}
 \begin{split}
  W(x,p) & = \int d^4y\,\sqrt{-g(x)}\,e^{-ip\cdot y/\hbar}\,\rho(x,y)\,, \\
  \rho(x,y) & = \langle\bar\psi(x,y/2) \otimes \psi(x,-y/2)\rangle
 \end{split}
\end{equation}
with $\psi(x)$ being the Dirac spinor on $\bbM$, $\bar\psi(x)\equiv \psi^\dag(x)\gamma^\hzero$, $\bar\psi\ola{O}\equiv [O \psi]^\dag \gamma^\hzero$ for an operator $O$, $\psi(x,y) =  \exp(y\cdot D)\psi(x)$, $\bar \psi(x,y) = \bar\psi(x)\exp(y\cdot\ola{D})$, and $[\bar\psi\otimes \psi]_{ab}=\bar\psi_b \psi_a$ ($a,b=1-4$).
Note that $D_\mu$ acting on the Dirac spinor involves $A_\mu$ to keep the $\text{U(1)}$ gauge covariance:
\begin{equation}
 D_\mu\psi(x,y)
 = \Bigl(
 		\nabla_\mu
 		- \Gamma^\lambda_{\mu\nu} y^\nu\partial^y_\lambda
 		+ iA_\mu/\hbar
   \Bigr) \psi(x,y) \,,
\end{equation}
where we recall that $\nabla_\mu\psi$ further involves the spin connection, i.e., $\nabla_\mu\psi=(\partial_\mu + \Gamma_\mu)\psi$. 
In Minkowski spacetime, Eq.~\eqref{eq:W} is reduced to a simple form with the Wilson line~\cite{Vasak:1987um}:
$\rho(x,y)=\langle\bar\psi(x_+)\otimes {\cal P}\exp[\tfrac{-i}{\hbar}\int_{x_-}^{x_+}dz\cdot A(z)] \psi(x_-)\rangle$ with $x_\pm=x\pm y/2$ and $\cal P$ the path ordering symbol.
In this paper, we focus on the collisionless fermions, so the spinor field obeys the Dirac equation
\begin{equation}
 \label{eq:diraceq}
 \gamma^\mu(\nabla_\mu + iA_\mu/\hbar)\,\psi(x)
 = \bar{\psi}(x)\,({\ola\nabla}_\mu - iA_\mu/\hbar)\gamma^\mu
 = 0\,.
\end{equation}

Computing $D_\mu\rho(x,y)$ and $\partial_\mu^y\rho(x,y)$ with the help of the Dirac equation~\eqref{eq:diraceq}, we derive:
\begin{equation}
 \label{eq:Weq-full}
 \begin{split}
  & \gamma^\mu \biggl(p_\mu + \frac{i\hbar}{2}D_\mu\biggr)\, W
  = i\hbar\gamma^\mu\int d^4y\,\sqrt{-g(x)}\,e^{-ip\cdot y/\hbar} \\
  &\qquad\times\Bigl\langle
  \bar\psi(x,y/2) \otimes (-\calH_\mu + \calG_\mu\bigr)\psi(x,-y/2) \\
  &\qquad\qquad\qquad\qquad -\bar\psi(x,y/2)\lcalG_\mu\otimes\psi(x,-y/2)
   \Bigr\rangle\,,\\
  &\ \quad \calH_\mu \psi(x,y)
  = -\frac{iy^\nu}{\hbar}\sum_{n=0}^\infty\frac{\bigl[\calC(y\cdot D)\bigr]^n}{(n+1)!}
  \,G_{\mu\nu}\,\psi(x,y) \,,\\
  &\ \quad \calG_\mu \psi(x,y)
  = -\frac{iy^\nu}{2\hbar}\sum_{n=0}^\infty\frac{\bigl[\calC(y\cdot D)\bigr]^n}{(n+2)!}
  \, G_{\mu\nu}\,\psi(x,y)\,,
 \end{split}
\end{equation}
where $\calC(X)Y \equiv[X,Y]$ represents a commutator. The details of the derivation of Eq.~\eqref{eq:Weq-full} can be found in Appendix~\ref{app:transport}; see also Ref.~\cite{Fonarev:1993ht}.
We defined $G_{\mu\nu} \equiv -i\hbar[D_\mu,D_\nu]$ as the total curvature tensor on $\bbT\bbM$ or $\bbT^*\bbM$.
For instance, we have $G_{\mu\nu}\,\psi(x,y) = (F_{\mu\nu} +\frac{\hbar}{4}R_{\mu\nu\alpha\beta}\sigma^{\alpha\beta} -i\hbar{R^\rho}_{\sigma\mu\nu}y^\sigma\partial^y_\rho) \psi(x,y)$, where the Riemann tensor is ${R^\rho}_{\sigma\mu\nu} = 2\partial_{[\nu}\Gamma_{\mu]\sigma}^\rho +2\Gamma_{\lambda[\nu}^\rho\Gamma^\lambda_{\mu]\sigma}$ with $X_{[\mu}Y_{\nu]} = \frac{1}{2}(X_\mu Y_\nu - Y_\nu X_\mu)$.
The transport equation~\eqref{eq:Weq-full} involves the full quantum correction coupled with electromagnetic field and curved background.
In Minkowski spacetime, Eq.~\eqref{eq:Weq-full} reproduces the transport equation derived in Ref.~\cite{Vasak:1987um}.

In practice, Eq.~\eqref{eq:Weq-full} is a powerful tool for the semiclassical analysis with the systematic expansion in terms of $\hbar$.
Let us adopt the power counting scheme with $p_\mu = O(1)$ and $y^\mu \sim i\hbar\partial^\mu_p = O(\hbar)$.
After a lengthy but straightforward calculation (see Appendix~\ref{app:semiclassical}), the transport equation for the Wigner function up to $O(\hbar^2)$ is written down, as follows:%
~\footnote{We actually keep the $O(\hbar^3)$ terms, which are necessary to derive Eqs.~\eqref{eq:R1}-\eqref{eq:R3} at $O(\hbar^2)$.}
\begin{equation}
 \begin{split}
  \label{eq:Weq-2nd}
   & \gamma^\mu\biggl(\Pi_\mu + \frac{i\hbar}{2} \Delta_\mu
   \biggr)W  \\
   & = \frac{i\hbar^2}{32}\gamma^\mu
   		\biggl (
   			 R_{\mu\nu\alpha\beta}
  		+ \frac{i\hbar}{6}\partial_p\cdot\nabla R_{\mu\nu\alpha\beta}
  		\biggr)
    		  \partial_p^\nu \Bigl[ W,\; \sigma^{\alpha\beta}\Bigr] \,,
 \end{split}
\end{equation}
with
\begin{equation}
 \begin{split}
 \label{eq:Pi_Delta}
 \Pi_\mu
  &= p_\mu
   	- \frac{\hbar^2}{12}(\nabla_\rho F_{\mu\nu})\partial^\nu_p\partial^\rho_p
  	+ \frac{\hbar^2}{24}{R^\rho}_{\sigma\mu\nu}\partial^\sigma_p\partial_p^\nu p_\rho
  	  +\frac{\hbar^2}{4}R_{\mu\nu}\partial_p^\nu\,, \\
 \Delta_\mu
  & = \nabla_\mu +\bigl(-F_{\mu\lambda}
  	 + \Gamma_{\mu\lambda}^\nu p_\nu\bigr)\partial_p^\lambda
  	 - \frac{\hbar^2}{12}(\nabla_\rho R_{\mu\nu})\partial_p^\rho\partial_p^\nu \\
  &\quad  - \frac{\hbar^2}{24}(\nabla_\lambda{R^\rho}_{\sigma\mu\nu})
  	    \partial_p^\nu\partial_p^\sigma\partial_p^\lambda p_\rho
  	 + \frac{\hbar^2}{8}{R^\rho}_{\sigma\mu\nu}\partial_p^\nu\partial_p^\sigma D_\rho \\
  &\quad + \frac{\hbar^2}{24} ( \nabla_\alpha\nabla_\beta F_{\mu\nu}
  	  + 2{R^\rho}_{\alpha\mu\nu}F_{\beta\rho} )
  	    \partial_p^\nu \partial_p^\alpha\partial_p^\beta\,,
 \end{split}
\end{equation}
where $R_{\mu\nu}={R^\rho}_{\mu\rho\nu}$ is the Ricci tensor.
Further we decompose Eq.~\eqref{eq:Weq-2nd} with the basis of the Clifford algebra:
$W =\tfrac{1}{4}[\calF +\gamma^5\calP+\gamma^\mu\calV_\mu+\gamma^5\gamma^\mu \calA_\mu+\tfrac{1}{2}\sigma^{\mu\nu}{\calS}_{\mu\nu}]$.
Then we obtain
\begin{eqnarray}
\label{eq:R1}
 & \dis\Delta\cdot\calR
 = \frac{\hbar^2}{24}(\nabla_\rho R_{\mu\nu}) \partial_p^\rho\partial_p^\mu\calR^\nu\,,\\
\label{eq:R2}
 & \dis\Pi\cdot \calR
 =\frac{\hbar^2}{8}R_{\mu\nu}\partial_p^\mu \calR^\nu\,,\\
\label{eq:R3}
 & \dis{\hbar}\Delta_{[\mu} \calR_{\nu]}
-\varepsilon_{\mu\nu\rho\sigma}\Pi^\rho\calR^\sigma
=-\frac{\hbar^2}{16}\varepsilon_{\mu\nu\alpha\beta}R^{\alpha\beta\rho\sigma}\partial_\rho^p\calR_\sigma\,,
\end{eqnarray}
with $\calR_\mu = (\calV_\mu+\calA_\mu)/2$ (see Appendix~\ref{app:spinor} for the derivation).
The first equation will be the kinetic equation for right-handed Weyl fermions, while the second and third serve as constraints.
The equations for $\calL_\mu=(\calV_\mu-\calA_\mu)/2$ are the same, except for a sign change in front of the first term of Eq.~\eqref{eq:R3}.


\section{Chiral kinetic equation at $O(\hbar)$}
Now we focus on the kinetic equation for $\calR^\mu(x,p)$ at $O(\hbar)$.
Equations~\eqref{eq:R1}-\eqref{eq:R3} are reduced to
\begin{eqnarray}
 \label{eq:eomI}
 & \Delta\cdot\calR = 0 \,, \\
 \label{eq:eomII}
 & p\cdot\calR = 0\,, \\
 \label{eq:eomIII}
 & \hbar \Delta_{[\mu} \calR_{\nu]}
  - \varepsilon_{\mu\nu\rho\sigma} p^\rho \calR^\sigma = 0
\end{eqnarray}
with $\Delta_\mu = \nabla_\mu +\bigl(-F_{\mu\lambda} + \Gamma_{\mu\lambda}^\nu p_\nu\bigr)\partial_p^\lambda$.
Thanks to the horizontal-lift prescription, we can solve Eqs.~\eqref{eq:eomI}-\eqref{eq:eomIII} in the same manner as that in flat spacetime.
The general solution is given by~\cite{Hidaka:2016yjf,Huang:2018wdl}
\begin{equation}
 \label{eq:Rmu}
 \calR^\mu
 = 4\pi\delta(p^2)\biggl[
 p^\mu -\frac{\hbar}{p^2}\tF^{\mu\nu}p_\nu
 + \hbar\Sigma_n^{\mu\nu}\Delta_\nu
 \biggr]f + O(\hbar^2)
\end{equation}
with $\tF^{\mu\nu} = \varepsilon^{\mu\nu\rho\sigma}F_{\rho\sigma}/2$ and $f=f(x,p)$ being the distribution function.
The last term is called the side-jump term~\cite{Chen:2014cla};
we introduced the spin tensor $\Sigma_n^{\mu\nu}\equiv \varepsilon^{\mu\nu\lambda\rho} p_\lambda n_\rho/ (2\,p\cdot n)$, where $n^\mu(x)$ is an arbitrary vector to satisfy $n\cdot p\neq 0$ and $n^2=1$.
This vector field accounts for an ambiguity in defining the spin for massless particles~\cite{Chen:2015gta}.
Different $n^\mu$'s correspond to different spin-frames and they are connected via $n'^\mu = {L^\mu}_\nu n^\nu = e^{\;\;\mu}_\halpha{\Lambda^\halpha}_\hbeta e^\hbeta_{\;\;\nu} n^\nu$ with ${\Lambda^\halpha}_\hbeta$ being a matrix representation of the local Lorentz transformation.

Plugging Eq.~\eqref{eq:Rmu} into Eq.~\eqref{eq:eomI}, we eventually obtain
\begin{equation}
 \begin{split}
 \label{eq:keq1st-n}
 &\delta\bigl(p^2-\hbar F_{\alpha\beta}\Sigma_n^{\alpha\beta}\bigr)\biggl[
 p\cdot\Delta+\hbar\biggl(\frac{n_\mu \tF^{\mu\nu}}{p\cdot n}
  + \Delta_\mu\Sigma^{\mu\nu}_n
 \biggr)\Delta_\nu \\
 &\qquad\qquad
 + \frac{\hbar}{2} \Sigma^{\mu\nu}_n \bigl(\nabla_\rho F_{\mu\nu}
 - p_\lambda {R^\lambda}_{\rho\mu\nu}\bigr) \partial_p^\rho\biggr] f = 0 \,.
 \end{split}
\end{equation}
This is the curved-spacetime generalization of the conventional chiral kinetic equation~\cite{Hidaka:2016yjf,Huang:2018wdl}.
Several comments are in order.
(I)~In the classical limit $\hbar\rightarrow 0$ we reproduce the Einstein-Vlasov equation:
$\delta(p^2)\, p^\mu\bigl[\partial_\mu +(- F_{\mu\nu} + \Gamma_{\mu\nu}^\lambda p_\lambda)\partial^\nu_p\bigr] f = 0$.
(II)~The spin connection $\Gamma_\mu$ is unrelated to $\Sigma_n^{\mu\nu}$.
Indeed, since $\calR^\mu$ is a vector, such a connection can never appear in Eq.~\eqref{eq:eomI}.
(III)~The Riemann curvature naively seems to be an $O(\hbar^2)$ correction, as Eqs.~\eqref{eq:R1}-\eqref{eq:R3} show.
However, once coupled with the side-jump term, it emerges even at $O(\hbar)$ in Eq.~\eqref{eq:keq1st-n}.
This term represents the so-called spin-curvature force~\cite{Mathisson:1937zz,*mathisson2010republication,Papapetrou:1951pa} for chiral fermions.
(IV)~On the other hand, the curvature does not appear in the delta function, which designates the on-shell condition.
However, this would not be the case at $O(\hbar^2)$.
In fact, from the viewpoint of field theory, the dispersion relation (without U(1) gauge field) reads $p^2 - \hbar^2R/4 = 0$ due to $(-i\hbar\gamma^\mu \nabla_\mu)^2\psi = -\hbar^2(\nabla_\mu\nabla^\mu+R/4)\psi = 0$~\cite{parker2009quantum}
(see also Ref.~\cite{Flachi:2017vlp} for a curvature correction to the CVE).

\section{Equilibrium state}
To reveal the physical content of Eq.~\eqref{eq:keq1st-n}, we consider the equilibrium state. We drop $A_\mu$ for simplicity.
At equilibrium, $f$ is generally written as a function of the linear combination of the collisional conserved quantities, i.e., the particle number, the linear momentum, and the angular momentum.
Therefore we have $f= \feq(g)$ 
with $g=\alpha(x)+\beta^\mu(x)p_\m +\hbar\gamma_{\mu\nu}(x)\Sigma^{\mu\nu}_n$.
Note that the orbital angular momentum is involved in the second term.
Plugging $\feq(g)$ into Eq.~\eqref{eq:keq1st-n} and requiring it to hold for arbitrary $p_\mu$, we arrive at the following constraints:
\begin{eqnarray}
 \label{eq:beta}
 & \nabla_\mu \beta_\nu + \nabla_\nu \beta_\mu = g_{\mu\nu}\phi(x) \,,\\
 \label{eq:alpha-gamma}
 & \nabla_\mu \alpha = 0 \,,\quad
 \dis \gamma_{\mu\nu} = \frac{1}{2}\nabla^\perp_{[\mu} \beta_{\nu]} \,,
\end{eqnarray}
where $\phi$ is an arbitrary scalar function and $\perp$ represents the component perpendicular to $n^\mu$.
In Appendix~\ref{app:equilibrium} we present the derivation of Eq.~\eqref{eq:beta} and Eq.~\eqref{eq:alpha-gamma}.

We have three comments about the above equations.
(I)~Equation~\eqref{eq:beta} is the conformal Killing equation.
Choosing a timelike $\beta^\mu$, we define the fluid velocity and temperature via $\beta^\m=\b U^\mu$ (with $U^2=1$) and $T=1/\b$, respectively. The physical meaning of $\phi$ is the expansion rate of the fluid:
$\phi = \frac{1}{2}\nabla\cdot\beta$, which follows from Eq.~\eqref{eq:beta}.
Thus the fluid is kept equilibrium under such an expansion.
This is understood as the conformal invariance in the massless Dirac theory.
Note that for massive particles $\phi$ must vanish, as the expansion can drive the system out of equilibrium.
(II)~From Eq.~\eqref{eq:alpha-gamma}, we find that $\alpha$ is a constant scalar.
We define the chemical potential through $\alpha = -\beta\mu$.
(III)~The equilibrium distribution is eventually given by $f=\feq(g)$ with
\begin{equation}
 \label{eq:feq}
   g = \beta(-\mu + p\cdot U)
   + \frac{\hbar}{2}\Sigma^{\mu\nu}_n\nabla_{\mu} \bigl(\beta U_{\nu}\bigr)\,.
\end{equation}
The last term expresses the spin-vorticity coupling.

\section{Rotating coordinate}
As the first application, we use our framework to revisit the derivation of the CVE by considering a rotating coordinate.
Let us choose a constant angular velocity $\bomega = (\omega^1,\omega^2,\omega^3)$ and hereafter set $A_\mu=0$.
The corresponding metric tensor reads
\begin{equation}
 \label{eq:metric}
 g_{00} = 1 - \bu^2\,, \quad g_{0i} = u^i\,,\quad g_{ij} = -\delta_{ij}
\end{equation}
with $\bu = (u^1,u^2,u^3) = \bx\times\bomega$.
The nonzero components of the Christoffel symbol are $\Gamma_{00}^i = -x^i\bomega^2 +(\bx\cdot\bomega)\omega^i = (\bu\times\bomega)^i$ and $\Gamma_{0j}^i = \Gamma_{j0}^i = -\varepsilon^{ijk}\omega^k$ (with $\varepsilon^{123}=1$), which lead to ${R^\rho}_{\sigma\mu\nu} = 0$.
The metric has an infinite red-shift surface at distance $r=1/|\bomega|$ away from the rotating axis.
We focus on the spacetime region inside this surface, and thus ignore the boundary effect of the system.
Such an assumption works as long as the angular velocity is small enough compared with other characteristic scales of the system~\cite{Vilenkin:1979ui,Ebihara:2016fwa}.
Therefore in the following analysis of the CVE, we consider the slowly rotating coordinate with $|\bomega| \ll T$ or $|\bomega| \ll \mu$.

In this case, the metric~\eqref{eq:metric} admits two timelike Killing vectors;
$K^\mu_{\rm in}=g^{\mu0}$ and $K_{\rm rot}^\m=\d^\m_0$.
Note that the former (latter) corresponds to the inertial (rotating) observers%
~\footnote{To be more specific, the rotating (Minkowski) coordinate is considered as the coordinate chart of the rotating (inertial) observer.}.
The velocities of these two observers are
\begin{equation}
 \calUin^\mu =(1,\bu) \,,\quad
 \calUrot^\mu =(g_{00})^{-\frac{1}{2}}\delta_0^\mu \,,
\end{equation}
which are normalized as $\mathcal{U}^2 = 1$.
From the on-shell condition $p^2 = g^{\mu\nu}p_\mu p_\nu = 0$, we obtain
\begin{eqnarray}
 \label{eq:dispersion1}
 & \epsilon_\bp \equiv K_{\rm rot}^\mu p_\mu=p_0
 = |\bp|+\bu\cdot\bp\,, \\
 \label{eq:velocity1}
 & \dis \bv_\bp = \frac{\partial\epsilon_\bp}{\partial\bp} = \hbp + \bu \,,
\end{eqnarray}
where $\bv_\bp$ denotes the group velocity and the three-momentum is defined as $\bp = -(p_1, p_2, p_3)$.
Similarly, we can obtain ${\epsilon}^{\rm in}_\bp \equiv K_{\rm in}^\mu p_\mu=p^0 = |\bp|$ and $\bv^{\rm in}_\bp =\partial{\epsilon}^{\rm in}_\bp/\partial\bp = \hbp$. Thus $\bp$ is identified as the momentum observed by the inertial observer. Note that the second terms in Eqs.~\eqref{eq:dispersion1} and ~\eqref{eq:velocity1} correspond to the rotating energy and velocity shifts, respectively.
In the following, we analyze the CKT with several choices of $n^\mu$ and $U^\mu$.

\subsection{Inertial fluid}
First of all, we consider an inertial fluid (i.e., a fluid at rest in flat spacetime) with a rotating observer.
We set $U^\mu = n^\mu =\calUin^\mu$ and $K^\mu=K^\mu_{\rm rot}$.
Performing the $p_0$-integration of Eq.~\eqref{eq:keq1st-n}, we find (for the particle channel only; antiparticle channel is similar)
\begin{equation}
 \label{eq:VEeq}
 \biggl[
  \frac{\partial}{\partial t} + \bv_\bp\cdot\frac{\partial}{\partial\bx}
  +(\bp\times\bomega)\cdot\frac{\partial}{\partial\bp}
 \biggr] f(t,\bx,\bp) = 0
\end{equation}
with $f(t,\bx,\bp) = f(t,\bx,\bp,p_0=\epsilon_\bp)$.
From the above equation, we identify $\dot{\bx} = \bv_\bp$ and $\dot{\bp} = \bp\times\bx$, which reproduce the Coriolis and centrifugal force: $\ddot{\bx} = 2\dot{\bx}\times\bomega -\bomega\times(\bomega\times\bx)$.
From Eq.~\eqref{eq:Rmu}, the particle number current reads
\begin{eqnarray}
 \label{eq:Jmu}
  &\dis J^\mu = (J^0\,,\bJ)
    \equiv \int \frac{d^4 p}{(2\pi)^4\sqrt{-g(x)}} \calR^\mu \,,\\
 \label{eq:J_inertial_n}
  & \dis \bJ = \intp \biggl[\bv_\bp
   -\hbar|\bp|\bOmega_\bp\times\frac{\partial}{\partial\bx}\biggr] f(t,\bx,\bp)
\end{eqnarray}
with $\intp = \int d^3p\,(2\pi)^{-3}$ and $\bOmega_\bp = \hbp/(2|\bp|^2)$ being the Berry curvature.
Note that due to $\nabla_\mu \calUin^\nu = 0$, at equilibrium all the $O(\hbar)$ corrections disappear in Eq.~\eqref{eq:J_inertial_n}, and thus it is just the classical Liouville current: $\bJ = \intp \bv_\bp \feq$.
Also from Eqs.~\eqref{eq:Rmu} and~\eqref{eq:feq} for $U^\mu = \calUin^\mu$, we can check that the same is true for arbitrary $n^\mu$.
Therefore, the CVE is never induced by an inertial fluid, independently of the observer's reference frame and the spin-frame choosing vector $n^\mu$.

\subsection{Rotating fluid}
In this case we adopt $U^\mu = \calUrot^\mu$, that is, we consider a fluid at rest in the rotating coordinate.
Hereafter let us focus on the small $\omega$ limit to simplify the discussions.

First, we choose $n^\mu = \calUin^\mu$, which leads to the kinetic equation and the current as the same forms as Eq.~\eqref{eq:VEeq} and~\eqref{eq:J_inertial_n}, respectively.
However, physical quantities are affected by quantum corrections.
When we take $f=\feq(g) = 1/(e^g+1)$ and append the antiparticle contribution (for which $\mu$ is replaced with $-\mu$), the $O(\bomega)$ terms in Eq.~\eqref{eq:J_inertial_n} yield
\begin{equation}
 \label{eq:CVE}
 \bJ_\text{CVE}
 = \hbar\bomega\biggl(\frac{\mu^2}{4\pi^2} + \frac{T^2}{12}\biggr)\,,
\end{equation}
which is the well-known CVE current. The spin-vorticity coupling term in Eq.~\eqref{eq:feq} are prominent to induce $\bJ_\text{CVE}$.
Because of this coupling, the first term in Eq.~\eqref{eq:J_inertial_n} gives $1/3$ of $\bJ_\text{CVE}$, while the second yields $2/3$~\cite{Chen:2014cla,Huang:2018aly}.

Second, we employ $n^\mu = \calUrot^\mu$.
It is more convenient to work with a new three-momentum defined as $\bq = (p^1,p^2,p^3)$%
~\footnote{
This is the change of the phase space variables that yields a nontrivial Jacobian which calls for $\partial_\mu\to\partial_\mu + (\partial_\mu p^\nu)\partial_\nu^p$.
}
, whose physical meaning will be explained later.
After the $p_0$ integration, the kinetic equation reads
\begin{equation}
 \label{eq:keq2}
 \begin{split}
  & \Biggl[
  (1+2\hbar\,|\bq|\bomega\cdot\bOmega_\bq)\frac{\partial}{\partial t}
  + \Bigl\{\tbv_\bq
  + 2\hbar|\bq|(\tbv_\bq\cdot\bOmega_\bq)\, \bomega
  \Bigr\}\cdot\frac{\partial}{\partial\bx}  \\
  & \qquad
  + 2|\bq|(\tbv_\bq\times\bomega)\cdot\frac{\partial}{\partial\bq}
  \Biggr] f(t,\bx,\bq) = 0
 \end{split}
\end{equation}
with the modified velocity $\tbv_\bq = \partial\tilde{\epsilon}_\bq/\partial\bq$ and energy dispersion
\begin{equation}
 \label{eq:tilde_epsilon}
 \tilde \epsilon_\bq = |\bq| - \frac{\hbar}{2}\hbq\cdot\bomega \,.
\end{equation}

The above kinetic equation exhibits an analogy between magnetism and rotation under two types of the correspondence, i.e., $|\bq|\bomega \leftrightarrow \bB$ in $\tepsilon_\bq$ (and so in $\tbv_\bq$), and $2|\bq|\bomega \leftrightarrow \bB$ elsewhere, reflecting the fact that the Land\'e $g$ factor is $2$ for spin-1/2 particles.
In other words, the spin-vorticity coupling plays a role of the magnetization coupling, and the Coriolis force can be regarded as a fictitious Lorentz force. This suggests that $\bq$ is the momentum observed by the rotating observer.
Indeed, Eq.~\eqref{eq:tilde_epsilon} shows that the classical dispersion is linear to $|\bq|$. For this reason, Eq.~\eqref{eq:keq2} are represented only with quantities in the rotating coordinate.
We note that the factor in front of $\partial /\partial t$ in Eq.~\eqref{eq:keq2} represents the quantum modification to the phase space measure~\cite{Dayi:2016foz,Huang:2018aly}.

From Eq.~\eqref{eq:Rmu}, we compute the particle number current as
\begin{equation}
 \label{eq:J_rotating_n}
 \bJ
  = \int_\bq \biggl[
  \tbv_\bq + 2\hbar|\bq|(\tbv_\bq\cdot\bOmega_\bq)\,\bomega
 \biggr] f(t,\bx,\bq) + O(\omega^2)\,,
\end{equation}
which, once substituted with $f=\feq(g) = 1/(e^g+1)$, reproduces Eq.~\eqref{eq:CVE} again.
Note that the first term does not contribute to the CVE current.
In other words, the Coriolis force is responsible for generating the CVE whereas the spin-vorticity coupling is not.
This explains why the heuristic replacement $\bB\to 2|\bq|\bomega$ works correctly~\cite{Stephanov:2012ki} in computing the CVE current.

Some comments are in order.
(I)~The above analysis shows that the origin of the CVE can be interpreted differently for different $n^\mu$.
For the inertial (rotating) spin-frame vector $n^\mu=\calUin^\mu$ ($\calUrot^\mu$), the CVE is induced through the spin-vorticity coupling (the Coriolis force). This is a clear demonstration for the nature of spinning massless particles: the total angular momentum is frame-dependently decomposed into the spin and the orbital parts~\cite{Skagerstam:1992er,Chen:2015gta,Stone:2015kla}.
(II)~However, in both cases with $n^\mu = \calUin^\mu$ and $n^\mu = \calUrot^\mu$, we derive the same CVE current~\eqref{eq:CVE}.
Indeed, the choice of $n^\m$ is superficially irrelevant to the CVE, as it is compensated by the side-jump effect.
This is confirmed from the fact that for arbitrary $n^\mu$, the equilibrium current is derived as a spin-frame-independent form:
for $f=\feq(g)=\feq^\zero + \frac{\hbar}{2}\Sigma_n^{\mu\nu}\nabla_\mu\beta_\nu (d\feq^\zero/dg)+O(\hbar^2)$ with $\feq^\zero = f(g=-\beta\mu + p\cdot \beta)$, Eq.~\eqref{eq:Jmu} is reduced to
\begin{equation}
 \label{eq:nindepend}
 \begin{split}
   J^\mu_\text{eq}
   & = \int\frac{d^4p\,\, 2 \delta(p^2)}{(2\pi)^3\sqrt{-g(x)}}
    \biggl[ p^\mu
   -\hbar\beta\frac{\omega^\mu}{2} (p\cdot U) \frac{d}{dg}
   \biggr]\feq^\zero\,,
 \end{split}
\end{equation}
with $\omega^\mu=\frac{1}{2}\e^{\mu\nu\rho\sigma}U_\nu\nabla_{\rho}U_{\sigma}$.
At the same time, we note that Eq.~\eqref{eq:nindepend} also holds for arbitrary curved spacetime.
This explains why the CVE current~\eqref{eq:CVE} is the same as that in Minkowski coordinate~\cite{Chen:2015gta,Dayi:2018xdy,Gao:2018jsi}.
The CVE is hence intrinsic for rotating fluid, of which the velocity configuration satisfies $\omega^\mu  \neq 0$.
\\

\section{Summary and outlook}
We extended the framework of the chiral kinetic theory (CKT) to curved spacetime, based on quantum field theory.
The CKT in curved spacetime is a primary tool for non-equilibrium chiral dynamics under the general-relativistic effect.
This enables us to investigate the anomalous transport phenomena in various chiral matter systems with (effective) gravitational field or non-inertial forces, such as supernova or neutron star environment~\cite{Janka:2012wk,Yamamoto:2015gzz}, rotating/expanding quark-gluon plasma~\cite{Jiang:2016woz,Deng:2016gyh,Ebihara:2017suq}, thermal systems with the temperature gradient~\cite{Luttinger:1964zz,Mameda:pre}, and Weyl/Dirac semimetals under strain~\cite{PhysRevLett.115.177202,Cortijo:2016wnf,PhysRevX.6.041046} or possibly torsion~\cite{Sumiyoshi:2015eda}.

As an application, we analyzed the CKT in a rotating coordinate, and clarified the frame-dependent interpretation for the chiral vortical effect (CVE).
Our calculation showed that although the CVE receives contributions from both the spin-vorticity coupling and Coriolis force depending on the choice of the defining frame of spin, their sum is independent of both the observer's frame and the spin-frame.
In this paper, we did not discuss about the relation between the finite-temperature term in the CVE current and the gravitational anomaly~\cite{Landsteiner:2011cp}.
On the other hand, it is still left open if such a term is induced by the global anomaly~\cite{Golkar:2012kb,Golkar:2015oxw,Jensen:2012kj,Glorioso:2017lcn}.
The CKT in curved spacetime is an auspicious candidate to lead to a model-independent answer to this mystery.
This will be shown in a future publication.

\begin{acknowledgments}
We thank Omer~Faruk~Dayi, Tomoya~Hayata, Yoshimasa~Hidaka, Bei-Lok~Hu, Kristan~Jensen, Jinfeng~Liao, Qun~Wang, and Yi~Yin, for useful discussions and valuable comments.
This work is supported by the China Postdoctoral Science Foundation under grant No.~2017M621345 (K.~M.), and the Young 1000 Talents Program of China, NSFC through Grants Nos.~11535012 and~11675041 (X.-G.~H.).
\end{acknowledgments}

\appendix
\section{Horizontal lift}\label{app:horizontal}
We derive the proper covariant derivative for functions in the cotangent bundle $\bbT^*\bbM$ from the viewpoint of the parallel displacement~\cite{dewitt2011bryce}. First, suppose that $\Phi(x,p)$ is a scalar function on $\bbT^*\bbM$.
Under the infinitesimal diffeomorphism $x^\mu\to x'^\mu = x^\mu + dx^\mu$, the variation $\delta\Phi$ involves two parts: one comes from the $x$-dependence of $\Phi$ and the other from the $p$-dependence.
That is, we write
\begin{equation}
 \begin{split}
  \delta\Phi
   & = \delta_x\Phi + \partial^\mu_p \Phi \delta p_\mu \,,
 \end{split}
\end{equation}
where $\delta p_\mu$ is the variation of $p_\mu$ under this diffeomorphic transformation.
By definition, we have $\delta\Phi = 0$ and $\delta p_\mu = \Gamma_{\mu\nu}^\rho p_\rho dx^\nu$, which leads to
\begin{equation}
 \delta_x\Phi = -\Gamma_{\mu\nu}^\rho p_\rho dx^\nu \,.
\end{equation}
Now we define a derivative $D_\mu$ as
\begin{equation}
 \label{eq:DPhidx}
 D_\mu \Phi\,dx^\mu \equiv \Phi(x+dx,p)-\Bigl[\Phi(x,p) + \delta_x\Phi(x,p)\Bigr] \,.
\end{equation}
Thus we obtain
\begin{equation}
 D_\mu \Phi = \Bigl(\partial_\mu + \Gamma_{\mu\nu}^\rho p_\rho\partial_p^\nu \Bigr) \Phi \,,
\end{equation}
which is the horizontal lift of $\partial_\mu$ to the cotangent bundle for a scalar field.
In the same manner, the covariant derivative for an arbitrary tensor field in the $\bbT^*\bbM$ is derived as
\begin{equation}
 D_\mu = \nabla_\mu + \Gamma_{\mu\nu}^\rho p_\rho\partial_p^\nu \,,
\end{equation}
with $\nabla_\mu$ the usual covariant derivative on $\bbM$. Similarly, we can define the horizontal lift of $\nabla_\mu$ to the tangent bundle $\bbT\bbM$ as given in the main text.

\section{Identities of the Dirac matrices}
We present some identities of the Dirac matrices which are useful in the derivation of the chiral kinetic equations.
From the definition $ \{ \gamma^\mu,\gamma^\nu \} = 2g^{\mu\nu}$ and $\gamma^5 = (-i/4!)\varepsilon_{\mu\nu\rho\sigma} \gamma^\mu\gamma^\nu\gamma^\rho\gamma^\sigma$, we find
\begin{eqnarray}
 & \dis \gamma^5 \sigma^{\mu\nu} = \frac{i}{2}\varepsilon^{\mu\nu\rho\sigma} \sigma_{\rho\sigma}\,,\\
 \label{eq:3gammas:a}
 & \gamma^\mu\gamma^\nu\gamma^\rho
 = g^{\mu\nu}\gamma^\rho + g^{\nu\rho}\gamma^\mu  - g^{\rho\mu}\gamma^{\nu}
 - i\varepsilon^{\mu\nu\rho\sigma}\gamma^5\gamma_\sigma,
\end{eqnarray}
where $\sigma^{\mu\nu} = (i/2)[\gamma^\mu,\gamma^\nu]$.
From these relations, we can prove the following useful identities:
\begin{eqnarray}
 \label{eq:gamma1:a}
 & \gamma^\mu \gamma^\nu
  = g^{\mu\nu} -i\sigma^{\mu\nu} \,,\\
 \label{eq:gamma2:a}
 & \gamma^\mu \bigl[\gamma^\nu, \sigma^{\alpha\beta} \bigr]
  = -4i g^{\mu[\alpha}g^{\beta]\nu}
  	- 4 \sigma^{\mu[\alpha}g^{\beta]\nu} \,, \\
  \label{eq:gamma3:a}
 & \dis \gamma^\mu  \gamma^5 \gamma^\nu
  = -g^{\mu\nu} \gamma^5
  	- \frac{1}{2}\varepsilon^{\mu\nu\alpha\beta}\sigma_{\alpha\beta} \,, \\
 \label{eq:gamma4:a}
 & \dis \gamma^\mu \bigl[\gamma^5\gamma^\nu, \sigma^{\alpha\beta} \bigr]
  = 4i\gamma^5 g^{\mu[\alpha} g^{\beta]\nu}
  	+ 2i\varepsilon^{\rho\sigma \mu[\alpha}g^{\beta]\nu}\sigma_{\rho\sigma}.
\end{eqnarray}

\section{Transport equation for $W(x,p)$}\label{app:transport}
The Wigner function of Dirac fermions is defined by
\begin{equation}
 \label{eq:W-app:a}
 \begin{split}
  W(x,p) = \int_y \rho(x,y)\,,  \quad
  \rho(x,y) = \langle\bar\psi_+ \otimes \psi_-\rangle \,.
 \end{split}
\end{equation}
Here we have introduced the following notations:
$\int_y = \int d^4y\,\sqrt{-g(x)}\,e^{-ip\cdot y/\hbar}$, $\psi_- = \psi(x,-y/2) = e^{-y\cdot D/2}\psi(x)$ and $\bar\psi_+ = \bar\psi(x,y/2) = \bar\psi(x)e^{y\cdot \lD/2}$.
Note that $D_\mu$ acting on the Dirac spinor involves $A_\mu$ to keep the $\text{U(1)}$ gauge covariance, that is,
\begin{equation}
 D_\mu\psi(x,y)
 = \Bigl(
 	\nabla_\mu - \Gamma^\lambda_{\mu\nu} y^\nu\partial^y_\lambda + iA_\mu/\hbar
   \Bigr)\psi(x,y) \,,
\end{equation}
with $\nabla_\mu\psi(x)=(\partial_\mu + \Gamma_\mu)\psi(x)$ and $\Gamma_\mu$ the spin connection. We consider the free Dirac field operators that obey
\begin{equation}
 \label{eq:diraceq:a}
 \gamma^\mu(\nabla_\mu + iA_\mu/\hbar)\,\psi(x)
 = \bar{\psi}(x)\,({\ola\nabla}_\mu - iA_\mu/\hbar)\gamma^\mu
 = 0\,.
\end{equation}
We assume that the surface integral for the Wigner transformation vanishes, that is,
\begin{equation}
 \label{eq:total-der:a}
 \begin{split}
   0
   & = \int d^4 y\, \partial_\mu^y \Bigl[\sqrt{-g(x)}e^{-ip\cdot y/\hbar} \rho(x,y) \Bigr] \\
   & =  -\frac{ip_\mu}{\hbar} W
  		+ \int_y \partial_\mu^y \rho(x,y) \,.
 \end{split}
\end{equation}
In order to calculate the second term, we utilize
\begin{eqnarray}
 \label{eq:Dpsi:a}
  & D_\mu\psi(x,y) = e^{y\cdot D}D_\mu\psi(x) - \calH_\mu\psi(x,y) \,, \\
 \label{eq:delypsi:a}
  & \partial^y_\mu \psi(x,y) = D_\mu\psi(x,y) + 2\calG_\mu\psi(x,y) \,,
\end{eqnarray}
which are derived from the operator identity $e^{Y} X e^{-Y} = e^{\mathcal{C}(Y)}X $ with $\mathcal{C}(Y) Z \equiv [Y,Z]$.
Here $\calH_\mu$ are $\calG_\mu$ are defined as
\begin{eqnarray}
 \label{eq:H:a}
  & \dis \calH_\mu \psi(x,y)
  = -\frac{iy^\nu}{\hbar}\sum_{n=0}^\infty\frac{\bigl[\calC(y\cdot D)\bigr]^n}{(n+1)!}
  \,G_{\mu\nu}\,\psi(x,y) \,, \\
 \label{eq:G:a}
  & \dis \calG_\mu \psi(x,y)
  = -\frac{iy^\nu}{2\hbar}\sum_{n=0}^\infty\frac{\bigl[\calC(y\cdot D)\bigr]^n}{(n+2)!}
  \, G_{\mu\nu}\,\psi(x,y)\,.
\end{eqnarray}
Here $G_{\mu\nu} \equiv -i\hbar[D_\mu,D_\nu]$ is the total curvature tensor on $\bbT\bbM$ and $\bbT^*\bbM$, e.g.,
\begin{widetext}
\begin{eqnarray}
  & \dis G_{\mu\nu} \psi(x,y)
  = \Bigl[ H_{\mu\nu}
  -i\hbar{R^\rho}_{\sigma\mu\nu}y^\sigma\partial^y_\rho\Bigr] \psi(x,y) \,, \\
  & \dis H_{\mu\nu}
  	=  F_{\mu\nu} +\frac{\hbar}{4}R_{\mu\nu\alpha\beta}\sigma^{\alpha\beta}\,, \quad
  F_{\mu\nu}
  	= 2\partial_{[\mu} A_{\nu]}\,, \quad
  {R^\rho}_{\sigma\mu\nu}
  	= 2\bigl(
  		\partial_{[\nu} \Gamma^\rho_{\mu]\sigma}
  		+ \Gamma_{\lambda[\nu}^\rho \Gamma_{\mu]\sigma}^\lambda
  	   \bigr)
\end{eqnarray}
with $X^{[\mu}Y^{\nu]} = \frac{1}{2}(X^\mu Y^\nu - X^\nu Y^\mu)$.
Then we compute
\begin{eqnarray}
 \label{eq:delrho:a}
  \int_y \partial_\mu^y \rho(x,y)
   = \frac{1}{2}D_\mu W
  	  + \int_y
  	  		\Bigl\langle
  	  			 \bar\psi_+ \otimes (\calH_\mu -\calG_\mu)\psi_-
  	  			+ \bar\psi_+ \lcalG_\mu \otimes \psi_-
  	 		\Bigr\rangle
 	  - \int_y
  	  		\Bigl\langle
  	  			\bar\psi_+ \otimes e^{-y\cdot D/2} D_\mu \psi(x)
     		\Bigr\rangle \,.
\end{eqnarray}
Contracted with $\gamma^\mu$, the last term vanishes due to the Dirac equation~\eqref{eq:diraceq:a}, and thus we derive
\begin{eqnarray}
 \label{eq:Weq-full:a}
   \gamma^\mu \biggl(p_\mu + \frac{i\hbar}{2}D_\mu\biggr)\, W
    = i\hbar\gamma^\mu\int_y \Bigl\langle
  \bar\psi_+ \otimes (-\calH_\mu + \calG_\mu\bigr)\psi_-
   -\bar\psi_+ \lcalG_\mu\otimes\psi_-
   \Bigr\rangle\,.
\end{eqnarray}
This is Eq.~\eqref{eq:Weq-full} in the main text.
\end{widetext}

\section{Semiclassical expansion up to $O(\hbar^3)$}\label{app:semiclassical}
Now we perform the semiclassical expansion of the transport equation~\eqref{eq:Weq-full:a} (that is, Eq.~\eqref{eq:Weq-full} in the main text) with the power counting scheme
\begin{eqnarray}
 p_\mu = O(1)\,,\quad y^\mu \sim i\hbar\partial^\mu_p = O(\hbar).
\end{eqnarray}
It should be noticed that $D_\mu$ can lead to terms of $O(\hbar^{-1})$ only when acting on $\psi(x,y)$:
\begin{eqnarray}
 D_\mu \psi(x,y) = O(\hbar^{-1})\,.
\end{eqnarray}
Since $(y\cdot D)\psi(x,y)$ is the same type function as $\psi(x,y)$ (i.e., a diffeomorphism scalar, local-Lorentz spinor), we readily find
\begin{widetext}
\begin{eqnarray}
  \Bigl[\calC(y\cdot D)\Bigr]^n G_{\mu\nu} \psi(x,y)
  = \Bigl[\calC(y\cdot D)\Bigr]^n \Bigl[ H_{\mu\nu}
  -i\hbar{R^\rho}_{\sigma\mu\nu}y^\sigma\partial^y_\rho\Bigr] \psi(x,y) \,.
\end{eqnarray}
Then $[\calC(y\cdot D)]^n G_{\mu\nu} \psi(x,y)$ is calculated as follows:
\begin{eqnarray}
  \dis \calC(y\cdot D)G_{\mu\nu} \psi(x,y)
  &=&
 	\Bigl[
 		y\cdot D
 		\bigl(
 			H_{\mu\nu}
 			-i\hbar {R^\rho}_{\sigma\mu\nu}y^\sigma\partial^y_\rho
 		\bigr)
 		+ i \hbar {R^\rho}_{\sigma\mu\nu} y^\sigma D_\rho
	\Bigr]  \psi(x,y)  \,, \\
 \dis [\calC(y\cdot D)]^2 G_{\mu\nu} \psi(x,y)
 &=& \Bigl[
   (y\cdot D)^2 F_{\mu\nu}
   + 2 i\hbar (y\cdot D){R^\rho}_{\sigma\mu\nu} y^\sigma D_\rho
   - y^\alpha y^\sigma {R^\rho}_{\sigma\mu\nu}F_{\alpha\rho}
   \Bigr] \psi(x,y) + O(\hbar^3) \,.
\end{eqnarray}
Note that $[\calC(y\cdot D)]^n G_{\mu\nu} \psi(x,y)$ for $n\geq 3$ does not yield any $O(\hbar^2)$ contribution.
Thus we obtain
\begin{eqnarray}
 \label{eq:RH}
  & & \dis i\hbar\int_y
  	\Bigl\langle
  		\bar\psi_+ \otimes (-\calH_\mu + \calG_\mu\bigr)\psi_-
  		-\bar\psi_+ \lcalG_\mu \otimes \psi_-
  	\Bigr\rangle  \nonumber \\
  &=&  \frac{i\hbar}{8}
  		\Bigl[
  			3 H_{\mu\nu}\partial_p^\nu W
  			+ W \lpart_p^\nu H_{\mu\nu}
  		\Bigr]
  	  + \frac{\hbar^2}{48}
  	  	\Bigl[
  	  		5 \partial_p\cdot \nabla H_{\mu\nu} \partial_p^\nu W
  			- W \lpart_p^\nu \lpart_p\cdot \nabla H_{\mu\nu}
  		\Bigr] \nonumber \\
  & & \dis
  	  - \frac{i\hbar^3}{48}
  		 \Bigl[
  			\nabla_\alpha\nabla_\beta F_{\mu\nu} \partial_p^\alpha\partial_p^\beta
   			- {R^\rho}_{\sigma\mu\nu}F_{\alpha\rho}
   				\partial_p^\alpha \partial_p^\sigma
   		\Bigr] \partial_p^\nu W
  	+ \frac{\hbar^2}{8} {R^\rho}_{\sigma\mu\nu}\partial_p^\nu
  			( 3 {\calX^\sigma}_\rho - {\calY^\sigma}_\rho )
  	 - \frac{i\hbar^3}{48} \partial_p\cdot\nabla
  	 	{R^\rho}_{\sigma\mu\nu}\partial_p^\nu
   			( 5 {\calX^\sigma}_\rho + {\calY^\sigma}_\rho ) \nonumber \\
  & & \dis
  	+ \frac{i\hbar^3}{48} {R^\rho}_{\sigma\mu\nu}
  		\partial_p^\nu \partial_p^\sigma
  		 ( 5 \calZ_\rho + \calW_\rho )
  	 + \frac{\hbar^4}{192} \partial_p\cdot \nabla {R^\rho}_{\sigma\mu\nu}
  		 \partial_p^\nu \partial_p^\sigma
  		 	( 7 \calZ_\rho - \calW_\rho)
  	 + O(\hbar^4) \,,
\end{eqnarray}
In the above equation we have defined
\begin{eqnarray}
 & \dis {\calX^\mu}_\nu
  = \int_y y^\mu\Bigl\langle \bar\psi_+\otimes \partial^y_\nu\psi_- \Bigr\rangle \,, \quad
  {\calY^\mu}_\nu
  = \int_y y^\mu\Bigl\langle \bar\psi_+ \lpart^y_\nu \otimes \psi_-\Bigr\rangle \,, \\
 &\dis \calZ_\mu
  = \int_y \Bigl\langle \bar\psi_+\otimes D_\mu \psi_- \Bigr\rangle \,, \quad
 \calW_\mu
  = \int_y \Bigl\langle \bar\psi_+ \lD_\mu  \otimes \psi_-\Bigr\rangle \,.
\end{eqnarray}
Although each of these is not a simple expression, their combinations are reduced to
\begin{eqnarray}
  \label{eq:XY}
 & \dis {\calX^\sigma}_\rho - {\calY^\sigma}_\rho
   = -\frac{i\hbar}{2} \partial_p^\sigma D_\rho W
  		- \frac{i\hbar}{4} F_{\lambda\rho}\partial_p^\lambda\partial_p^\sigma W
  		+ O(\hbar^2) \,,\qquad
   {\calX^\sigma}_\rho + {\calY^\sigma}_\rho
   = -\partial^\sigma_p p_\rho W \,, \\
  \label{eq:ZW}
 & \dis \calZ_\rho - \calW_\rho
   = -\frac{2ip_\rho}{\hbar} W + O(\hbar) \,, \qquad
 \calZ_\rho + \calW_\rho
   = D_\rho W\,,
\end{eqnarray}
which follow from Eq.~\eqref{eq:delypsi:a}.
From these relations and Eq.~\eqref{eq:3gammas:a}, we eventually derive
\begin{equation}
   \label{eq:Weq-2nd:a}
  \gamma^\mu\biggl(\Pi_\mu + \frac{i\hbar}{2} \Delta_\mu \biggr) W
   =  \frac{i}{4}\gamma^\mu\Theta_{\mu\alpha\beta} \Bigl[ W,\; \sigma^{\alpha\beta}\Bigr]
\end{equation}
with
\begin{eqnarray}
  \dis \Pi_\mu
  &=& p_\mu
   	- \frac{\hbar^2}{12}(\nabla_\rho F_{\mu\nu})\partial^\n_p\partial^\rho_p
  	+ \frac{\hbar^2}{24}{R^\rho}_{\sigma\mu\nu}\partial^\sigma_p\partial_p^\nu p_\rho
  	  +\frac{\hbar^2}{4}R_{\mu\nu}\partial_p^\nu\,, \\
 \dis \Delta_\mu
  &=& \nabla_\mu +\bigl(-F_{\mu\lambda}
  	 + \Gamma_{\mu\lambda}^\nu p_\nu\bigr)\partial_p^\lambda
  	 - \frac{\hbar^2}{12}(\nabla_\r R_{\m\n})\partial_p^\r\partial_p^\nu
  	 - \frac{\hbar^2}{24}(\nabla_\l{R^\rho}_{\sigma\mu\nu})
  	    \partial_p^\nu\partial_p^\s\partial_p^\lambda p_\rho \nonumber \\
  && \dis
  	 + \frac{\hbar^2}{8}{R^\rho}_{\sigma\mu\nu}\partial_p^\nu\partial_p^\sigma D_\rho
  	 + \frac{\hbar^2}{24} ( \nabla_\alpha\nabla_\beta F_{\mu\nu}
  	 + 2{R^\rho}_{\alpha\mu\nu}F_{\beta\rho} )
  	    \partial_p^\nu \partial_p^\alpha\partial_p^\beta\,,  \\
 \dis \Theta_{\mu\alpha\beta}
   &=& \frac{\hbar^2}{8} R_{\mu\nu\alpha\beta} \partial_p^\nu
  		+ \frac{i\hbar^3}{48}\partial_p\cdot\nabla R_{\mu\nu\alpha\beta} \partial_p^\nu \,.
 \end{eqnarray}
They are Eqs.~\eqref{eq:Weq-2nd} and ~\eqref{eq:Pi_Delta} in the main text.
\end{widetext}

\section{Spinor decomposition}\label{app:spinor}
In the last step, we decompose Eq.~\eqref{eq:Weq-2nd:a} with the basis of the Clifford algebra.
The Wigner function $W$ is written as
\begin{equation}
 W = \frac{1}{4}
 		\biggl[
 			\calF + \gamma^5\calP
 			+ \gamma\cdot\calV + \gamma^5\gamma\cdot\calA
 			+ \frac{1}{2}\sigma^{\mu\nu}{\calS}_{\mu\nu}
 		\biggr] \,.
\end{equation}
The coefficients $\calV_\mu$ and $\calA_\mu$ correspond to the vector and axial currents, respectively:
$\text{tr}[\gamma^\mu W] = \calV^\mu$ and $\text{tr}[\gamma^\mu \gamma^5 W] = \calA^\mu$.
This means that the right- and left-handed currents are represented as
\begin{equation}
 \calR_\mu = \frac{1}{2}(\calV_\mu+\calA_\mu)\,,\quad
 \calL_\mu = \frac{1}{2}(\calV_\mu-\calA_\mu)\,.
\end{equation}
Due to the masslessness in Eq.~\eqref{eq:Weq-2nd:a}, the vector and axial channels are decoupled from the scalar, pseudo-scalar, and tensor ones. Focusing on the vector and axial channels, we find
\begin{eqnarray}
 \label{eq:Weq-VA:a}
  \dis \gamma\cdot\Lambda (\gamma\cdot\calV + \gamma^5 \gamma\cdot\calA)
     = \frac{i}{4}\gamma^\mu \Theta_{\mu\alpha\beta}
    		\Bigl[
    			\gamma\cdot\calV + \gamma^5 \gamma \cdot\calA,
    			\sigma^{\alpha\beta}
    		\Bigr]\,, \nonumber\\
\end{eqnarray}
where we have introduced the shorthand notation $\Lambda_\mu = \Pi_\mu + ({i\hbar}/2) \Delta_\mu$.
By inserting Eqs.~\eqref{eq:gamma1:a}-\eqref{eq:gamma4:a} into Eq.~\eqref{eq:Weq-VA:a} and extracting three parts proportional to $1$, $\gamma^5$ and $\sigma^{\mu\nu}$, respectively, we obtain
\begin{eqnarray}
   & (\Lambda_\nu - {\Theta^{\mu}}_{\mu\nu}) \calV^\nu = 0 \,, \\
   & (\Lambda_\nu - {\Theta^{\mu}}_{\mu\nu}) \calA^\nu = 0 \,, \\
   & \dis
    \Lambda_{[\mu} \calV_{\nu]} - \Theta_{[\mu\nu]\rho}\calV^\rho
    - \frac{i}{2} \varepsilon_{\mu\nu\rho\sigma} (
    \Lambda^\rho\calA^\sigma
    - \Theta^{\rho\sigma\lambda}\calA_\lambda  )
    = 0 \,.\qquad
\end{eqnarray}
Note that, by contracting $(i/2)\varepsilon^{\mu\nu}_{\;\;\;\;\alpha\beta}$ with the third equation, we obtain the same equation with the replacement $\calV^\mu\leftrightarrow\calA^\mu$.
Separating the real and imaginary parts, we finally derive
\begin{eqnarray}
 \label{eq:R1:a}
 & \dis \Delta\cdot\calR
  = \frac{\hbar^2}{24}\partial_p\cdot \nabla R_{\mu\nu}\partial_p^\mu \calR^\nu \,, \\
 \label{eq:R2:a}
 & \dis \Pi\cdot\calR = \frac{\hbar^2}{8}R_{\mu\nu}\partial_p^\mu \calR^\nu \,, \\
 \label{eq:R3:a}
 & \dis \hbar\Delta_{[\mu}\calR_{\nu]}
 - \varepsilon_{\mu\nu\rho\sigma}\Pi^{\rho}\calR^\sigma
 = -\frac{\hbar^2}{16}\varepsilon_{\mu\nu\rho\sigma}
  R^{\alpha\beta\rho\sigma}\partial_\alpha^p\calR_\beta \,,
\end{eqnarray}
where we used $2R_{\alpha[\mu\nu]\beta} = - R_{\alpha\beta\mu\nu}$ which follows from the Bianchi identity for the Riemann curvature.
These are Eqs.~\eqref{eq:R1}-~\eqref{eq:R3} in the main text.

\section{Equilibrium distribution}\label{app:equilibrium}
In this Appendix, we derive the general form of the equilibrium distribution in the absence of the electromagnetic field, that is,
$A_\mu = 0$. The kinetic equation at $O(\hbar)$ reads
\begin{equation}
 \delta(p^2)
 	\Bigl[
 		p\cdot D + \hbar(D_\mu \Sigma_n^{\mu\nu})D_\nu
 		- \frac{\hbar}{2}\Sigma_n^{\mu\nu}p^\lambda R_{\lambda\rho\mu\nu}\partial_p^\rho
 	\Bigr] f = 0 \,.
\end{equation}
At equilibrium, the distribution $f$ should be a function of the
linear combination of the collisional conserved quantities, namely, the particle number, the linear momentum, and
the angular momentum. Therefore, we parametrize the equilibrium distribution as
\begin{equation}
 f = \feq(g)\,,\quad
 g =\alpha(x) + \beta(x)\cdot p + \hbar\gamma^{\mu\nu}(x)\Sigma_{\mu\nu}^n \,,
\end{equation}
with $\alpha(x)=\alpha_0(x)+\hbar\alpha_1(x)$, $\beta^\mu(x)=\beta^\mu_0(x)+\hbar\beta^\mu_1(x)$, and $\gamma^{\mu\nu}(x)=\gamma^{\mu\nu}_0(x)+\hbar\gamma^{\mu\nu}_1(x)$ and $O(\hbar^2)$ terms omitted. Note that $n_\mu \gamma^{\mu\nu}=n_\nu \gamma^{\mu\nu}=0$. Then the kinetic equation yields
\begin{equation}
 \label{eq:eq-0th}
 \begin{split}
   & \delta(p^2)
 \Bigl(
 	p\cdot\nabla \alpha_0 + p^\mu p^\nu \nabla_\mu \beta_{0\nu}
 \Bigr) = 0 \,,
 \end{split}
\end{equation}
and
\begin{equation}
 \label{eq:eq-1st}
  \begin{split}
  & \delta(p^2)
 \Bigl[p\cdot\nabla \alpha_1 + p^\mu p^\nu \nabla_\mu \beta_{1\nu}+
 	(D_\mu\Sigma_n^{\mu\nu})(\nabla_\nu \alpha_0 + p^\lambda\nabla_\nu\beta_{0\lambda})\\
 & \qquad
	+ p\cdot D (\gamma_0^{\mu\nu}\Sigma_{\mu\nu}^n)
	- \frac{1}{2} \Sigma_n^{\mu\nu} R_{\lambda\rho\mu\nu}p^\lambda\beta_0^\rho
 \Bigr]= 0 \,.
 \end{split}
\end{equation}
From Eq.~\eqref{eq:eq-0th}, we find
\begin{equation}
\label{constraint0}
 \nabla_\mu \alpha_0 = 0 \,,\quad
 \nabla_\mu\beta_{0\nu} + \nabla_\nu\beta_{0\mu} = \phi_0(x) g_{\mu\nu} \,,
\end{equation}
where $\phi_0$ is an arbitrary scalar function.
Combined these constraints, the third term in Eq.~\eqref{eq:eq-1st} is calculated as
\begin{equation}
 \begin{split}
  & \delta(p^2)(D_\mu\Sigma_n^{\mu\nu})(\nabla_\nu \alpha_0 + p^\lambda\nabla_\nu\beta_{0\lambda}) \\
  & = \delta(p^2)
  		\Bigl[
  			D_\mu(\Sigma_n^{\mu[\nu} p^{\lambda]}\nabla_{\nu}\beta_{0\lambda})
  	  	- \Sigma_n^{\mu\nu} p^{\lambda}\nabla_\mu \nabla_{[\nu}\beta_{0\lambda]}
  	  	\Bigr] \\
  & = \delta(p^2)
  		\biggl[
  			-\frac{1}{2}p\cdot D
  				(\Sigma_n^{\mu\nu} \nabla_{\mu}\beta_{0\nu})
			+ \frac{1}{2} \Sigma_n^{\mu\nu} R_{\lambda\rho\mu\nu}p^\lambda\beta_0^\rho
  		\biggr] \,,
 \end{split}
\end{equation}
where we have used $ p_\mu \varepsilon_{\nu\rho\sigma\lambda} + p_\nu \varepsilon_{\rho\sigma\lambda\mu} + p_\rho \varepsilon_{\sigma\lambda\mu\nu} + p_\sigma \varepsilon_{\lambda\mu\nu\rho} + p_\lambda \varepsilon_{\mu\nu\rho\sigma} = 0 $, $\nabla_\mu\nabla_{[\nu}\beta_{0\lambda]} = -\beta_0^\alpha R_{\alpha\mu\nu\lambda} - g_{\mu[\nu}\nabla_{\lambda]} \phi_0$, $2R_{\alpha[\mu\nu]\beta} = - R_{\alpha\beta\mu\nu}$, and $\varepsilon^{\alpha\mu\nu\rho}R_{\beta\mu\nu\rho}=0$.
Therefore, Eq.~\eqref{eq:eq-1st} is reduced to
\begin{equation}
 \begin{split}
 & \delta(p^2)
  		\Bigl[
  			p\cdot\nabla \alpha_1 
  			+ p^\mu p^\nu \nabla_\mu \beta_{1\nu}
  			+ p\cdot D ( \Sigma_n^{\mu\nu}M_{\mu\nu} )		
  		\Bigr] = 0\,,
 \end{split}	
\end{equation}
where $M_{\m\n}=\gamma_{0\mu\nu}-\frac{1}{2} \nabla^\perp_{[\mu}\beta_{0\nu]}$ with $\perp$ here representing the component perpendicular to $n^\mu$ and we have used the fact that $\Sigma_n^{\mu\nu}\nabla_{\mu}\beta_{0\nu} = \Sigma_n^{\mu\nu}\nabla^\perp_{\mu}\beta_{0\nu}$. Therefore, we have
\begin{equation}
 \begin{split}
 & p\cdot\nabla \alpha_1 
 	+ p^\mu p^\nu \nabla_\mu \beta_{1\nu}
    + p\cdot D(\Sigma_n^{\mu\nu}M_{\mu\nu}) 
  = \sum_{l=1}^\infty \frac{\phi_l(x)}{2}(p^2)^l\,,
 \end{split}	
\end{equation}
where $\phi_l$ with $l\geq 1$ are arbitrary functions. Making a scale transformation, $p_\mu \rightarrow \lambda p_\mu$, and comparing different powers of $\lambda$, we obtain
\begin{eqnarray}
 \label{betafirst}
 &\nabla_\mu\beta_{1\nu} + \nabla_\nu\beta_{1\mu} = \phi_1(x) g_{\mu\nu},\\
  \label{phihigher}
 & \phi_l =0, \;\;\;\; l\geq2, \\
 \label{alphafirst}
  & p\cdot\nabla \alpha_1 + p\cdot D(
\Sigma_n^{\mu\nu}M_{\mu\nu})=0\,.
\end{eqnarray}
To proceed, we decompose $p_\mu$ with respect to $n_\mu$ as follows, $p_\mu=p_\parallel n_\mu +p^\perp_\mu$. Substituting into Eq.~\eqref{alphafirst} we find that the following conditions
\begin{equation}
\label{constraint1}
 \begin{split}
 & \nabla_\mu \alpha_1=0,\;\; M_{\mu\nu}=0 
 \end{split}	
\end{equation}
fulfill Eq.~\eqref{alphafirst} for arbitrary $p^\perp_\mu$. However, in case that $n_\mu$ is a constant, $\nabla_\mu n_\nu= 0$, the conditions to fulfill Eq.~\eqref{alphafirst} is
\begin{equation}
\label{constraint2}
 \begin{split}
 & n\cdot\nabla \alpha_1=0,\;\; \nabla_\lambda^\perp M_{\mu\nu}=0\;,\\
 & \nabla^\perp_\m\alpha_1 +\frac{1}{2}\varepsilon_{\rho\sigma\mu\nu}n^\nu n\cdot\nabla M^{\rho\sigma}=0\;\,,
 \end{split}	
\end{equation}
which nevertheless contains constraint~\eqref{constraint1} as a special case. Collecting Eq.~\eqref{constraint0}, Eq.~\eqref{betafirst}, and Eq.~\eqref{constraint1}, we obtain the equilibrium condition as given in  Eq.~\eqref{eq:beta} and Eq.~\eqref{eq:alpha-gamma} in the main text up to $O(\hbar)$. We also notice that in the case of $\nabla_\mu n_\nu=0$, the equilibrium state can maintain a difference between $\gamma_{0\mu\nu}$ and the thermal vorticity $\frac{1}{2} \nabla^\perp_{[\mu}\beta_{0\nu]}$ if a finite gradient of the chemical potential is present, as shown in the third equation in Eq.~\eqref{constraint2}, whose physical consequence deserves a more careful exploration in future.

\bibliography{ckt}

\begin{thebibliography}{68}%
\makeatletter
\providecommand \@ifxundefined [1]{%
 \@ifx{#1\undefined}
}%
\providecommand \@ifnum [1]{%
 \ifnum #1\expandafter \@firstoftwo
 \else \expandafter \@secondoftwo
 \fi
}%
\providecommand \@ifx [1]{%
 \ifx #1\expandafter \@firstoftwo
 \else \expandafter \@secondoftwo
 \fi
}%
\providecommand \natexlab [1]{#1}%
\providecommand \enquote  [1]{``#1''}%
\providecommand \bibnamefont  [1]{#1}%
\providecommand \bibfnamefont [1]{#1}%
\providecommand \citenamefont [1]{#1}%
\providecommand \href@noop [0]{\@secondoftwo}%
\providecommand \href [0]{\begingroup \@sanitize@url \@href}%
\providecommand \@href[1]{\@@startlink{#1}\@@href}%
\providecommand \@@href[1]{\endgroup#1\@@endlink}%
\providecommand \@sanitize@url [0]{\catcode `\\12\catcode `\$12\catcode
  `\&12\catcode `\#12\catcode `\^12\catcode `\_12\catcode `\%12\relax}%
\providecommand \@@startlink[1]{}%
\providecommand \@@endlink[0]{}%
\providecommand \url  [0]{\begingroup\@sanitize@url \@url }%
\providecommand \@url [1]{\endgroup\@href {#1}{\urlprefix }}%
\providecommand \urlprefix  [0]{URL }%
\providecommand \Eprint [0]{\href }%
\providecommand \doibase [0]{http://dx.doi.org/}%
\providecommand \selectlanguage [0]{\@gobble}%
\providecommand \bibinfo  [0]{\@secondoftwo}%
\providecommand \bibfield  [0]{\@secondoftwo}%
\providecommand \translation [1]{[#1]}%
\providecommand \BibitemOpen [0]{}%
\providecommand \bibitemStop [0]{}%
\providecommand \bibitemNoStop [0]{.\EOS\space}%
\providecommand \EOS [0]{\spacefactor3000\relax}%
\providecommand \BibitemShut  [1]{\csname bibitem#1\endcsname}%
\let\auto@bib@innerbib\@empty
\bibitem [{\citenamefont {Kharzeev}\ \emph {et~al.}(2008)\citenamefont
  {Kharzeev}, \citenamefont {McLerran},\ and\ \citenamefont
  {Warringa}}]{Kharzeev:2007jp}%
  \BibitemOpen
  \bibfield  {author} {\bibinfo {author} {\bibfnamefont {D.~E.}\ \bibnamefont
  {Kharzeev}}, \bibinfo {author} {\bibfnamefont {L.~D.}\ \bibnamefont
  {McLerran}}, \ and\ \bibinfo {author} {\bibfnamefont {H.~J.}\ \bibnamefont
  {Warringa}},\ }\href {\doibase 10.1016/j.nuclphysa.2008.02.298} {\bibfield
  {journal} {\bibinfo  {journal} {Nucl. Phys.}\ }\textbf {\bibinfo {volume}
  {A803}},\ \bibinfo {pages} {227} (\bibinfo {year} {2008})},\ \Eprint
  {http://arxiv.org/abs/0711.0950} {arXiv:0711.0950 [hep-ph]} \BibitemShut
  {NoStop}%
\bibitem [{\citenamefont {Fukushima}\ \emph {et~al.}(2008)\citenamefont
  {Fukushima}, \citenamefont {Kharzeev},\ and\ \citenamefont
  {Warringa}}]{Fukushima:2008xe}%
  \BibitemOpen
  \bibfield  {author} {\bibinfo {author} {\bibfnamefont {K.}~\bibnamefont
  {Fukushima}}, \bibinfo {author} {\bibfnamefont {D.~E.}\ \bibnamefont
  {Kharzeev}}, \ and\ \bibinfo {author} {\bibfnamefont {H.~J.}\ \bibnamefont
  {Warringa}},\ }\href {\doibase 10.1103/PhysRevD.78.074033} {\bibfield
  {journal} {\bibinfo  {journal} {Phys. Rev.}\ }\textbf {\bibinfo {volume}
  {D78}},\ \bibinfo {pages} {074033} (\bibinfo {year} {2008})},\ \Eprint
  {http://arxiv.org/abs/0808.3382} {arXiv:0808.3382 [hep-ph]} \BibitemShut
  {NoStop}%
\bibitem [{\citenamefont {Vilenkin}(1979)}]{Vilenkin:1979ui}%
  \BibitemOpen
  \bibfield  {author} {\bibinfo {author} {\bibfnamefont {A.}~\bibnamefont
  {Vilenkin}},\ }\href {\doibase 10.1103/PhysRevD.20.1807} {\bibfield
  {journal} {\bibinfo  {journal} {Phys. Rev.}\ }\textbf {\bibinfo {volume}
  {D20}},\ \bibinfo {pages} {1807} (\bibinfo {year} {1979})}\BibitemShut
  {NoStop}%
\bibitem [{\citenamefont {Erdmenger}\ \emph {et~al.}(2009)\citenamefont
  {Erdmenger}, \citenamefont {Haack}, \citenamefont {Kaminski},\ and\
  \citenamefont {Yarom}}]{Erdmenger:2008rm}%
  \BibitemOpen
  \bibfield  {author} {\bibinfo {author} {\bibfnamefont {J.}~\bibnamefont
  {Erdmenger}}, \bibinfo {author} {\bibfnamefont {M.}~\bibnamefont {Haack}},
  \bibinfo {author} {\bibfnamefont {M.}~\bibnamefont {Kaminski}}, \ and\
  \bibinfo {author} {\bibfnamefont {A.}~\bibnamefont {Yarom}},\ }\href
  {\doibase 10.1088/1126-6708/2009/01/055} {\bibfield  {journal} {\bibinfo
  {journal} {JHEP}\ }\textbf {\bibinfo {volume} {01}},\ \bibinfo {pages} {055}
  (\bibinfo {year} {2009})},\ \Eprint {http://arxiv.org/abs/0809.2488}
  {arXiv:0809.2488 [hep-th]} \BibitemShut {NoStop}%
\bibitem [{\citenamefont {Banerjee}\ \emph {et~al.}(2011)\citenamefont
  {Banerjee}, \citenamefont {Bhattacharya}, \citenamefont {Bhattacharyya},
  \citenamefont {Dutta}, \citenamefont {Loganayagam},\ and\ \citenamefont
  {Surowka}}]{Banerjee:2008th}%
  \BibitemOpen
  \bibfield  {author} {\bibinfo {author} {\bibfnamefont {N.}~\bibnamefont
  {Banerjee}}, \bibinfo {author} {\bibfnamefont {J.}~\bibnamefont
  {Bhattacharya}}, \bibinfo {author} {\bibfnamefont {S.}~\bibnamefont
  {Bhattacharyya}}, \bibinfo {author} {\bibfnamefont {S.}~\bibnamefont
  {Dutta}}, \bibinfo {author} {\bibfnamefont {R.}~\bibnamefont {Loganayagam}},
  \ and\ \bibinfo {author} {\bibfnamefont {P.}~\bibnamefont {Surowka}},\ }\href
  {\doibase 10.1007/JHEP01(2011)094} {\bibfield  {journal} {\bibinfo  {journal}
  {JHEP}\ }\textbf {\bibinfo {volume} {01}},\ \bibinfo {pages} {094} (\bibinfo
  {year} {2011})},\ \Eprint {http://arxiv.org/abs/0809.2596} {arXiv:0809.2596
  [hep-th]} \BibitemShut {NoStop}%
\bibitem [{\citenamefont {Son}\ and\ \citenamefont
  {Surowka}(2009)}]{Son:2009tf}%
  \BibitemOpen
  \bibfield  {author} {\bibinfo {author} {\bibfnamefont {D.~T.}\ \bibnamefont
  {Son}}\ and\ \bibinfo {author} {\bibfnamefont {P.}~\bibnamefont {Surowka}},\
  }\href {\doibase 10.1103/PhysRevLett.103.191601} {\bibfield  {journal}
  {\bibinfo  {journal} {Phys. Rev. Lett.}\ }\textbf {\bibinfo {volume} {103}},\
  \bibinfo {pages} {191601} (\bibinfo {year} {2009})},\ \Eprint
  {http://arxiv.org/abs/0906.5044} {arXiv:0906.5044 [hep-th]} \BibitemShut
  {NoStop}%
\bibitem [{\citenamefont {Kharzeev}\ \emph {et~al.}(2016)\citenamefont
  {Kharzeev}, \citenamefont {Liao}, \citenamefont {Voloshin},\ and\
  \citenamefont {Wang}}]{Kharzeev:2015znc}%
  \BibitemOpen
  \bibfield  {author} {\bibinfo {author} {\bibfnamefont {D.~E.}\ \bibnamefont
  {Kharzeev}}, \bibinfo {author} {\bibfnamefont {J.}~\bibnamefont {Liao}},
  \bibinfo {author} {\bibfnamefont {S.~A.}\ \bibnamefont {Voloshin}}, \ and\
  \bibinfo {author} {\bibfnamefont {G.}~\bibnamefont {Wang}},\ }\href {\doibase
  10.1016/j.ppnp.2016.01.001} {\bibfield  {journal} {\bibinfo  {journal} {Prog.
  Part. Nucl. Phys.}\ }\textbf {\bibinfo {volume} {88}},\ \bibinfo {pages} {1}
  (\bibinfo {year} {2016})},\ \Eprint {http://arxiv.org/abs/1511.04050}
  {arXiv:1511.04050 [hep-ph]} \BibitemShut {NoStop}%
\bibitem [{\citenamefont {Huang}(2016{\natexlab{a}})}]{Huang:2015oca}%
  \BibitemOpen
  \bibfield  {author} {\bibinfo {author} {\bibfnamefont {X.-G.}\ \bibnamefont
  {Huang}},\ }\href {\doibase 10.1088/0034-4885/79/7/076302} {\bibfield
  {journal} {\bibinfo  {journal} {Rept. Prog. Phys.}\ }\textbf {\bibinfo
  {volume} {79}},\ \bibinfo {pages} {076302} (\bibinfo {year}
  {2016}{\natexlab{a}})},\ \Eprint {http://arxiv.org/abs/1509.04073}
  {arXiv:1509.04073 [nucl-th]} \BibitemShut {NoStop}%
\bibitem [{\citenamefont {Hattori}\ and\ \citenamefont
  {Huang}(2017)}]{Hattori:2016emy}%
  \BibitemOpen
  \bibfield  {author} {\bibinfo {author} {\bibfnamefont {K.}~\bibnamefont
  {Hattori}}\ and\ \bibinfo {author} {\bibfnamefont {X.-G.}\ \bibnamefont
  {Huang}},\ }\href {\doibase 10.1007/s41365-016-0178-3} {\bibfield  {journal}
  {\bibinfo  {journal} {Nucl. Sci. Tech.}\ }\textbf {\bibinfo {volume} {28}},\
  \bibinfo {pages} {26} (\bibinfo {year} {2017})},\ \Eprint
  {http://arxiv.org/abs/1609.00747} {arXiv:1609.00747 [nucl-th]} \BibitemShut
  {NoStop}%
\bibitem [{\citenamefont {Koch}\ \emph {et~al.}(2017)\citenamefont {Koch},
  \citenamefont {Schlichting}, \citenamefont {Skokov}, \citenamefont
  {Sorensen}, \citenamefont {Thomas}, \citenamefont {Voloshin}, \citenamefont
  {Wang},\ and\ \citenamefont {Yee}}]{Skokov:2016yrj}%
  \BibitemOpen
  \bibfield  {author} {\bibinfo {author} {\bibfnamefont {V.}~\bibnamefont
  {Koch}}, \bibinfo {author} {\bibfnamefont {S.}~\bibnamefont {Schlichting}},
  \bibinfo {author} {\bibfnamefont {V.}~\bibnamefont {Skokov}}, \bibinfo
  {author} {\bibfnamefont {P.}~\bibnamefont {Sorensen}}, \bibinfo {author}
  {\bibfnamefont {J.}~\bibnamefont {Thomas}}, \bibinfo {author} {\bibfnamefont
  {S.}~\bibnamefont {Voloshin}}, \bibinfo {author} {\bibfnamefont
  {G.}~\bibnamefont {Wang}}, \ and\ \bibinfo {author} {\bibfnamefont {H.-U.}\
  \bibnamefont {Yee}},\ }\href {\doibase 10.1088/1674-1137/41/7/072001}
  {\bibfield  {journal} {\bibinfo  {journal} {Chin. Phys.}\ }\textbf {\bibinfo
  {volume} {C41}},\ \bibinfo {pages} {072001} (\bibinfo {year} {2017})},\
  \Eprint {http://arxiv.org/abs/1608.00982} {arXiv:1608.00982 [nucl-th]}
  \BibitemShut {NoStop}%
\bibitem [{\citenamefont {Charbonneau}\ and\ \citenamefont
  {Zhitnitsky}(2010)}]{Charbonneau:2009ax}%
  \BibitemOpen
  \bibfield  {author} {\bibinfo {author} {\bibfnamefont {J.}~\bibnamefont
  {Charbonneau}}\ and\ \bibinfo {author} {\bibfnamefont {A.}~\bibnamefont
  {Zhitnitsky}},\ }\href {\doibase 10.1088/1475-7516/2010/08/010} {\bibfield
  {journal} {\bibinfo  {journal} {JCAP}\ }\textbf {\bibinfo {volume} {1008}},\
  \bibinfo {pages} {010} (\bibinfo {year} {2010})},\ \Eprint
  {http://arxiv.org/abs/0903.4450} {arXiv:0903.4450 [astro-ph.HE]} \BibitemShut
  {NoStop}%
\bibitem [{\citenamefont {Ohnishi}\ and\ \citenamefont
  {Yamamoto}(2014)}]{Ohnishi:2014uea}%
  \BibitemOpen
  \bibfield  {author} {\bibinfo {author} {\bibfnamefont {A.}~\bibnamefont
  {Ohnishi}}\ and\ \bibinfo {author} {\bibfnamefont {N.}~\bibnamefont
  {Yamamoto}},\ }\href@noop {} {\  (\bibinfo {year} {2014})},\ \Eprint
  {http://arxiv.org/abs/1402.4760} {arXiv:1402.4760 [astro-ph.HE]} \BibitemShut
  {NoStop}%
\bibitem [{\citenamefont {Masada}\ \emph {et~al.}(2018)\citenamefont {Masada},
  \citenamefont {Kotake}, \citenamefont {Takiwaki},\ and\ \citenamefont
  {Yamamoto}}]{Masada:2018swb}%
  \BibitemOpen
  \bibfield  {author} {\bibinfo {author} {\bibfnamefont {Y.}~\bibnamefont
  {Masada}}, \bibinfo {author} {\bibfnamefont {K.}~\bibnamefont {Kotake}},
  \bibinfo {author} {\bibfnamefont {T.}~\bibnamefont {Takiwaki}}, \ and\
  \bibinfo {author} {\bibfnamefont {N.}~\bibnamefont {Yamamoto}},\ }\href
  {\doibase 10.1103/PhysRevD.98.083018} {\bibfield  {journal} {\bibinfo
  {journal} {Phys. Rev.}\ }\textbf {\bibinfo {volume} {D98}},\ \bibinfo {pages}
  {083018} (\bibinfo {year} {2018})},\ \Eprint
  {http://arxiv.org/abs/1805.10419} {arXiv:1805.10419 [astro-ph.HE]}
  \BibitemShut {NoStop}%
\bibitem [{\citenamefont {Miransky}\ and\ \citenamefont
  {Shovkovy}(2015)}]{Miransky:2015ava}%
  \BibitemOpen
  \bibfield  {author} {\bibinfo {author} {\bibfnamefont {V.~A.}\ \bibnamefont
  {Miransky}}\ and\ \bibinfo {author} {\bibfnamefont {I.~A.}\ \bibnamefont
  {Shovkovy}},\ }\href {\doibase 10.1016/j.physrep.2015.02.003} {\bibfield
  {journal} {\bibinfo  {journal} {Phys. Rept.}\ }\textbf {\bibinfo {volume}
  {576}},\ \bibinfo {pages} {1} (\bibinfo {year} {2015})},\ \Eprint
  {http://arxiv.org/abs/1503.00732} {arXiv:1503.00732 [hep-ph]} \BibitemShut
  {NoStop}%
\bibitem [{\citenamefont {Armitage}\ \emph {et~al.}(2018)\citenamefont
  {Armitage}, \citenamefont {Mele},\ and\ \citenamefont
  {Vishwanath}}]{Armitage:2017cjs}%
  \BibitemOpen
  \bibfield  {author} {\bibinfo {author} {\bibfnamefont {N.~P.}\ \bibnamefont
  {Armitage}}, \bibinfo {author} {\bibfnamefont {E.~J.}\ \bibnamefont {Mele}},
  \ and\ \bibinfo {author} {\bibfnamefont {A.}~\bibnamefont {Vishwanath}},\
  }\href {\doibase 10.1103/RevModPhys.90.015001} {\bibfield  {journal}
  {\bibinfo  {journal} {Rev. Mod. Phys.}\ }\textbf {\bibinfo {volume} {90}},\
  \bibinfo {pages} {015001} (\bibinfo {year} {2018})},\ \Eprint
  {http://arxiv.org/abs/1705.01111} {arXiv:1705.01111 [cond-mat.str-el]}
  \BibitemShut {NoStop}%
\bibitem [{\citenamefont {Burkov}(2015)}]{Burkov:2015hba}%
  \BibitemOpen
  \bibfield  {author} {\bibinfo {author} {\bibfnamefont {A.~A.}\ \bibnamefont
  {Burkov}},\ }\href {\doibase 10.1088/0953-8984/27/11/113201} {\bibfield
  {journal} {\bibinfo  {journal} {J. Phys. Condens. Matter}\ }\textbf {\bibinfo
  {volume} {27}},\ \bibinfo {pages} {113201} (\bibinfo {year} {2015})},\
  \Eprint {http://arxiv.org/abs/1502.07609} {arXiv:1502.07609
  [cond-mat.mes-hall]} \BibitemShut {NoStop}%
\bibitem [{\citenamefont {Stephanov}\ and\ \citenamefont
  {Yin}(2012)}]{Stephanov:2012ki}%
  \BibitemOpen
  \bibfield  {author} {\bibinfo {author} {\bibfnamefont {M.~A.}\ \bibnamefont
  {Stephanov}}\ and\ \bibinfo {author} {\bibfnamefont {Y.}~\bibnamefont
  {Yin}},\ }\href {\doibase 10.1103/PhysRevLett.109.162001} {\bibfield
  {journal} {\bibinfo  {journal} {Phys. Rev. Lett.}\ }\textbf {\bibinfo
  {volume} {109}},\ \bibinfo {pages} {162001} (\bibinfo {year} {2012})},\
  \Eprint {http://arxiv.org/abs/1207.0747} {arXiv:1207.0747 [hep-th]}
  \BibitemShut {NoStop}%
\bibitem [{\citenamefont {Son}\ and\ \citenamefont
  {Yamamoto}(2013)}]{Son:2012zy}%
  \BibitemOpen
  \bibfield  {author} {\bibinfo {author} {\bibfnamefont {D.~T.}\ \bibnamefont
  {Son}}\ and\ \bibinfo {author} {\bibfnamefont {N.}~\bibnamefont {Yamamoto}},\
  }\href {\doibase 10.1103/PhysRevD.87.085016} {\bibfield  {journal} {\bibinfo
  {journal} {Phys. Rev.}\ }\textbf {\bibinfo {volume} {D87}},\ \bibinfo {pages}
  {085016} (\bibinfo {year} {2013})},\ \Eprint {http://arxiv.org/abs/1210.8158}
  {arXiv:1210.8158 [hep-th]} \BibitemShut {NoStop}%
\bibitem [{\citenamefont {Gao}\ \emph {et~al.}(2012)\citenamefont {Gao},
  \citenamefont {Liang}, \citenamefont {Pu}, \citenamefont {Wang},\ and\
  \citenamefont {Wang}}]{Gao:2012ix}%
  \BibitemOpen
  \bibfield  {author} {\bibinfo {author} {\bibfnamefont {J.-H.}\ \bibnamefont
  {Gao}}, \bibinfo {author} {\bibfnamefont {Z.-T.}\ \bibnamefont {Liang}},
  \bibinfo {author} {\bibfnamefont {S.}~\bibnamefont {Pu}}, \bibinfo {author}
  {\bibfnamefont {Q.}~\bibnamefont {Wang}}, \ and\ \bibinfo {author}
  {\bibfnamefont {X.-N.}\ \bibnamefont {Wang}},\ }\href {\doibase
  10.1103/PhysRevLett.109.232301} {\bibfield  {journal} {\bibinfo  {journal}
  {Phys. Rev. Lett.}\ }\textbf {\bibinfo {volume} {109}},\ \bibinfo {pages}
  {232301} (\bibinfo {year} {2012})},\ \Eprint {http://arxiv.org/abs/1203.0725}
  {arXiv:1203.0725 [hep-ph]} \BibitemShut {NoStop}%
\bibitem [{\citenamefont {Chen}\ \emph {et~al.}(2013)\citenamefont {Chen},
  \citenamefont {Pu}, \citenamefont {Wang},\ and\ \citenamefont
  {Wang}}]{Chen:2012ca}%
  \BibitemOpen
  \bibfield  {author} {\bibinfo {author} {\bibfnamefont {J.-W.}\ \bibnamefont
  {Chen}}, \bibinfo {author} {\bibfnamefont {S.}~\bibnamefont {Pu}}, \bibinfo
  {author} {\bibfnamefont {Q.}~\bibnamefont {Wang}}, \ and\ \bibinfo {author}
  {\bibfnamefont {X.-N.}\ \bibnamefont {Wang}},\ }\href {\doibase
  10.1103/PhysRevLett.110.262301} {\bibfield  {journal} {\bibinfo  {journal}
  {Phys. Rev. Lett.}\ }\textbf {\bibinfo {volume} {110}},\ \bibinfo {pages}
  {262301} (\bibinfo {year} {2013})},\ \Eprint {http://arxiv.org/abs/1210.8312}
  {arXiv:1210.8312 [hep-th]} \BibitemShut {NoStop}%
\bibitem [{\citenamefont {Huang}(2016{\natexlab{b}})}]{Huang:2015mga}%
  \BibitemOpen
  \bibfield  {author} {\bibinfo {author} {\bibfnamefont {X.-G.}\ \bibnamefont
  {Huang}},\ }\href {\doibase 10.1038/srep20601} {\bibfield  {journal}
  {\bibinfo  {journal} {Sci. Rep.}\ }\textbf {\bibinfo {volume} {6}},\ \bibinfo
  {pages} {20601} (\bibinfo {year} {2016}{\natexlab{b}})},\ \Eprint
  {http://arxiv.org/abs/1506.03590} {arXiv:1506.03590 [cond-mat.quant-gas]}
  \BibitemShut {NoStop}%
\bibitem [{\citenamefont {Hidaka}\ \emph {et~al.}(2017)\citenamefont {Hidaka},
  \citenamefont {Pu},\ and\ \citenamefont {Yang}}]{Hidaka:2016yjf}%
  \BibitemOpen
  \bibfield  {author} {\bibinfo {author} {\bibfnamefont {Y.}~\bibnamefont
  {Hidaka}}, \bibinfo {author} {\bibfnamefont {S.}~\bibnamefont {Pu}}, \ and\
  \bibinfo {author} {\bibfnamefont {D.-L.}\ \bibnamefont {Yang}},\ }\href
  {\doibase 10.1103/PhysRevD.95.091901} {\bibfield  {journal} {\bibinfo
  {journal} {Phys. Rev.}\ }\textbf {\bibinfo {volume} {D95}},\ \bibinfo {pages}
  {091901} (\bibinfo {year} {2017})},\ \Eprint
  {http://arxiv.org/abs/1612.04630} {arXiv:1612.04630 [hep-th]} \BibitemShut
  {NoStop}%
\bibitem [{\citenamefont {Huang}\ \emph {et~al.}(2018)\citenamefont {Huang},
  \citenamefont {Shi}, \citenamefont {Jiang}, \citenamefont {Liao},\ and\
  \citenamefont {Zhuang}}]{Huang:2018wdl}%
  \BibitemOpen
  \bibfield  {author} {\bibinfo {author} {\bibfnamefont {A.}~\bibnamefont
  {Huang}}, \bibinfo {author} {\bibfnamefont {S.}~\bibnamefont {Shi}}, \bibinfo
  {author} {\bibfnamefont {Y.}~\bibnamefont {Jiang}}, \bibinfo {author}
  {\bibfnamefont {J.}~\bibnamefont {Liao}}, \ and\ \bibinfo {author}
  {\bibfnamefont {P.}~\bibnamefont {Zhuang}},\ }\href {\doibase
  10.1103/PhysRevD.98.036010} {\bibfield  {journal} {\bibinfo  {journal} {Phys.
  Rev.}\ }\textbf {\bibinfo {volume} {D98}},\ \bibinfo {pages} {036010}
  (\bibinfo {year} {2018})},\ \Eprint {http://arxiv.org/abs/1801.03640}
  {arXiv:1801.03640 [hep-th]} \BibitemShut {NoStop}%
\bibitem [{\citenamefont {Mueller}\ and\ \citenamefont
  {Venugopalan}(2017)}]{Mueller:2017arw}%
  \BibitemOpen
  \bibfield  {author} {\bibinfo {author} {\bibfnamefont {N.}~\bibnamefont
  {Mueller}}\ and\ \bibinfo {author} {\bibfnamefont {R.}~\bibnamefont
  {Venugopalan}},\ }\href {\doibase 10.1103/PhysRevD.96.016023} {\bibfield
  {journal} {\bibinfo  {journal} {Phys. Rev.}\ }\textbf {\bibinfo {volume}
  {D96}},\ \bibinfo {pages} {016023} (\bibinfo {year} {2017})},\ \Eprint
  {http://arxiv.org/abs/1702.01233} {arXiv:1702.01233 [hep-ph]} \BibitemShut
  {NoStop}%
\bibitem [{\citenamefont {Berry}(1984)}]{Berry:1984jv}%
  \BibitemOpen
  \bibfield  {author} {\bibinfo {author} {\bibfnamefont {M.~V.}\ \bibnamefont
  {Berry}},\ }\href {\doibase 10.1098/rspa.1984.0023} {\bibfield  {journal}
  {\bibinfo  {journal} {Proc. Roy. Soc. Lond.}\ }\textbf {\bibinfo {volume}
  {A392}},\ \bibinfo {pages} {45} (\bibinfo {year} {1984})}\BibitemShut
  {NoStop}%
\bibitem [{\citenamefont {Chen}\ \emph {et~al.}(2014)\citenamefont {Chen},
  \citenamefont {Son}, \citenamefont {Stephanov}, \citenamefont {Yee},\ and\
  \citenamefont {Yin}}]{Chen:2014cla}%
  \BibitemOpen
  \bibfield  {author} {\bibinfo {author} {\bibfnamefont {J.-Y.}\ \bibnamefont
  {Chen}}, \bibinfo {author} {\bibfnamefont {D.~T.}\ \bibnamefont {Son}},
  \bibinfo {author} {\bibfnamefont {M.~A.}\ \bibnamefont {Stephanov}}, \bibinfo
  {author} {\bibfnamefont {H.-U.}\ \bibnamefont {Yee}}, \ and\ \bibinfo
  {author} {\bibfnamefont {Y.}~\bibnamefont {Yin}},\ }\href {\doibase
  10.1103/PhysRevLett.113.182302} {\bibfield  {journal} {\bibinfo  {journal}
  {Phys. Rev. Lett.}\ }\textbf {\bibinfo {volume} {113}},\ \bibinfo {pages}
  {182302} (\bibinfo {year} {2014})},\ \Eprint {http://arxiv.org/abs/1404.5963}
  {arXiv:1404.5963 [hep-th]} \BibitemShut {NoStop}%
\bibitem [{\citenamefont {Chen}\ \emph {et~al.}(2015)\citenamefont {Chen},
  \citenamefont {Son},\ and\ \citenamefont {Stephanov}}]{Chen:2015gta}%
  \BibitemOpen
  \bibfield  {author} {\bibinfo {author} {\bibfnamefont {J.-Y.}\ \bibnamefont
  {Chen}}, \bibinfo {author} {\bibfnamefont {D.~T.}\ \bibnamefont {Son}}, \
  and\ \bibinfo {author} {\bibfnamefont {M.~A.}\ \bibnamefont {Stephanov}},\
  }\href {\doibase 10.1103/PhysRevLett.115.021601} {\bibfield  {journal}
  {\bibinfo  {journal} {Phys. Rev. Lett.}\ }\textbf {\bibinfo {volume} {115}},\
  \bibinfo {pages} {021601} (\bibinfo {year} {2015})},\ \Eprint
  {http://arxiv.org/abs/1502.06966} {arXiv:1502.06966 [hep-th]} \BibitemShut
  {NoStop}%
\bibitem [{\citenamefont {Gorbar}\ \emph {et~al.}(2017)\citenamefont {Gorbar},
  \citenamefont {Miransky}, \citenamefont {Shovkovy},\ and\ \citenamefont
  {Sukhachov}}]{Gorbar:2016ygi}%
  \BibitemOpen
  \bibfield  {author} {\bibinfo {author} {\bibfnamefont {E.~V.}\ \bibnamefont
  {Gorbar}}, \bibinfo {author} {\bibfnamefont {V.~A.}\ \bibnamefont
  {Miransky}}, \bibinfo {author} {\bibfnamefont {I.~A.}\ \bibnamefont
  {Shovkovy}}, \ and\ \bibinfo {author} {\bibfnamefont {P.~O.}\ \bibnamefont
  {Sukhachov}},\ }\href {\doibase 10.1103/PhysRevLett.118.127601} {\bibfield
  {journal} {\bibinfo  {journal} {Phys. Rev. Lett.}\ }\textbf {\bibinfo
  {volume} {118}},\ \bibinfo {pages} {127601} (\bibinfo {year} {2017})},\
  \Eprint {http://arxiv.org/abs/1610.07625} {arXiv:1610.07625
  [cond-mat.str-el]} \BibitemShut {NoStop}%
\bibitem [{\citenamefont {Carignano}\ \emph {et~al.}(2018)\citenamefont
  {Carignano}, \citenamefont {Manuel},\ and\ \citenamefont
  {Torres-Rincon}}]{Carignano:2018gqt}%
  \BibitemOpen
  \bibfield  {author} {\bibinfo {author} {\bibfnamefont {S.}~\bibnamefont
  {Carignano}}, \bibinfo {author} {\bibfnamefont {C.}~\bibnamefont {Manuel}}, \
  and\ \bibinfo {author} {\bibfnamefont {J.~M.}\ \bibnamefont
  {Torres-Rincon}},\ }\href {\doibase 10.1103/PhysRevD.98.076005} {\bibfield
  {journal} {\bibinfo  {journal} {Phys. Rev.}\ }\textbf {\bibinfo {volume}
  {D98}},\ \bibinfo {pages} {076005} (\bibinfo {year} {2018})},\ \Eprint
  {http://arxiv.org/abs/1806.01684} {arXiv:1806.01684 [hep-ph]} \BibitemShut
  {NoStop}%
\bibitem [{\citenamefont {Hidaka}\ \emph {et~al.}(2018)\citenamefont {Hidaka},
  \citenamefont {Pu},\ and\ \citenamefont {Yang}}]{Hidaka:2017auj}%
  \BibitemOpen
  \bibfield  {author} {\bibinfo {author} {\bibfnamefont {Y.}~\bibnamefont
  {Hidaka}}, \bibinfo {author} {\bibfnamefont {S.}~\bibnamefont {Pu}}, \ and\
  \bibinfo {author} {\bibfnamefont {D.-L.}\ \bibnamefont {Yang}},\ }\href
  {\doibase 10.1103/PhysRevD.97.016004} {\bibfield  {journal} {\bibinfo
  {journal} {Phys. Rev.}\ }\textbf {\bibinfo {volume} {D97}},\ \bibinfo {pages}
  {016004} (\bibinfo {year} {2018})},\ \Eprint
  {http://arxiv.org/abs/1710.00278} {arXiv:1710.00278 [hep-th]} \BibitemShut
  {NoStop}%
\bibitem [{\citenamefont {Hidaka}\ and\ \citenamefont
  {Yang}(2018)}]{Hidaka:2018ekt}%
  \BibitemOpen
  \bibfield  {author} {\bibinfo {author} {\bibfnamefont {Y.}~\bibnamefont
  {Hidaka}}\ and\ \bibinfo {author} {\bibfnamefont {D.-L.}\ \bibnamefont
  {Yang}},\ }\href {\doibase 10.1103/PhysRevD.98.016012} {\bibfield  {journal}
  {\bibinfo  {journal} {Phys. Rev.}\ }\textbf {\bibinfo {volume} {D98}},\
  \bibinfo {pages} {016012} (\bibinfo {year} {2018})},\ \Eprint
  {http://arxiv.org/abs/1801.08253} {arXiv:1801.08253 [hep-th]} \BibitemShut
  {NoStop}%
\bibitem [{\citenamefont {Landsteiner}\ \emph {et~al.}(2011)\citenamefont
  {Landsteiner}, \citenamefont {Megias},\ and\ \citenamefont
  {Pena-Benitez}}]{Landsteiner:2011cp}%
  \BibitemOpen
  \bibfield  {author} {\bibinfo {author} {\bibfnamefont {K.}~\bibnamefont
  {Landsteiner}}, \bibinfo {author} {\bibfnamefont {E.}~\bibnamefont {Megias}},
  \ and\ \bibinfo {author} {\bibfnamefont {F.}~\bibnamefont {Pena-Benitez}},\
  }\href {\doibase 10.1103/PhysRevLett.107.021601} {\bibfield  {journal}
  {\bibinfo  {journal} {Phys. Rev. Lett.}\ }\textbf {\bibinfo {volume} {107}},\
  \bibinfo {pages} {021601} (\bibinfo {year} {2011})},\ \Eprint
  {http://arxiv.org/abs/1103.5006} {arXiv:1103.5006 [hep-ph]} \BibitemShut
  {NoStop}%
\bibitem [{\citenamefont {Golkar}\ and\ \citenamefont
  {Son}(2015)}]{Golkar:2012kb}%
  \BibitemOpen
  \bibfield  {author} {\bibinfo {author} {\bibfnamefont {S.}~\bibnamefont
  {Golkar}}\ and\ \bibinfo {author} {\bibfnamefont {D.~T.}\ \bibnamefont
  {Son}},\ }\href {\doibase 10.1007/JHEP02(2015)169} {\bibfield  {journal}
  {\bibinfo  {journal} {JHEP}\ }\textbf {\bibinfo {volume} {02}},\ \bibinfo
  {pages} {169} (\bibinfo {year} {2015})},\ \Eprint
  {http://arxiv.org/abs/1207.5806} {arXiv:1207.5806 [hep-th]} \BibitemShut
  {NoStop}%
\bibitem [{\citenamefont {Golkar}\ and\ \citenamefont
  {Sethi}(2016)}]{Golkar:2015oxw}%
  \BibitemOpen
  \bibfield  {author} {\bibinfo {author} {\bibfnamefont {S.}~\bibnamefont
  {Golkar}}\ and\ \bibinfo {author} {\bibfnamefont {S.}~\bibnamefont {Sethi}},\
  }\href {\doibase 10.1007/JHEP05(2016)105} {\bibfield  {journal} {\bibinfo
  {journal} {JHEP}\ }\textbf {\bibinfo {volume} {05}},\ \bibinfo {pages} {105}
  (\bibinfo {year} {2016})},\ \Eprint {http://arxiv.org/abs/1512.02607}
  {arXiv:1512.02607 [hep-th]} \BibitemShut {NoStop}%
\bibitem [{\citenamefont {Jensen}\ \emph {et~al.}(2013)\citenamefont {Jensen},
  \citenamefont {Loganayagam},\ and\ \citenamefont {Yarom}}]{Jensen:2012kj}%
  \BibitemOpen
  \bibfield  {author} {\bibinfo {author} {\bibfnamefont {K.}~\bibnamefont
  {Jensen}}, \bibinfo {author} {\bibfnamefont {R.}~\bibnamefont {Loganayagam}},
  \ and\ \bibinfo {author} {\bibfnamefont {A.}~\bibnamefont {Yarom}},\ }\href
  {\doibase 10.1007/JHEP02(2013)088} {\bibfield  {journal} {\bibinfo  {journal}
  {JHEP}\ }\textbf {\bibinfo {volume} {02}},\ \bibinfo {pages} {088} (\bibinfo
  {year} {2013})},\ \Eprint {http://arxiv.org/abs/1207.5824} {arXiv:1207.5824
  [hep-th]} \BibitemShut {NoStop}%
\bibitem [{\citenamefont {Glorioso}\ \emph {et~al.}(2019)\citenamefont
  {Glorioso}, \citenamefont {Liu},\ and\ \citenamefont
  {Rajagopal}}]{Glorioso:2017lcn}%
  \BibitemOpen
  \bibfield  {author} {\bibinfo {author} {\bibfnamefont {P.}~\bibnamefont
  {Glorioso}}, \bibinfo {author} {\bibfnamefont {H.}~\bibnamefont {Liu}}, \
  and\ \bibinfo {author} {\bibfnamefont {S.}~\bibnamefont {Rajagopal}},\ }\href
  {\doibase 10.1007/JHEP01(2019)043} {\bibfield  {journal} {\bibinfo  {journal}
  {JHEP}\ }\textbf {\bibinfo {volume} {01}},\ \bibinfo {pages} {043} (\bibinfo
  {year} {2019})},\ \Eprint {http://arxiv.org/abs/1710.03768} {arXiv:1710.03768
  [hep-th]} \BibitemShut {NoStop}%
\bibitem [{\citenamefont {Basar}\ \emph {et~al.}(2013)\citenamefont {Basar},
  \citenamefont {Kharzeev},\ and\ \citenamefont {Zahed}}]{Basar:2013qia}%
  \BibitemOpen
  \bibfield  {author} {\bibinfo {author} {\bibfnamefont {G.}~\bibnamefont
  {Basar}}, \bibinfo {author} {\bibfnamefont {D.~E.}\ \bibnamefont {Kharzeev}},
  \ and\ \bibinfo {author} {\bibfnamefont {I.}~\bibnamefont {Zahed}},\ }\href
  {\doibase 10.1103/PhysRevLett.111.161601} {\bibfield  {journal} {\bibinfo
  {journal} {Phys. Rev. Lett.}\ }\textbf {\bibinfo {volume} {111}},\ \bibinfo
  {pages} {161601} (\bibinfo {year} {2013})},\ \Eprint
  {http://arxiv.org/abs/1307.2234} {arXiv:1307.2234 [hep-th]} \BibitemShut
  {NoStop}%
\bibitem [{\citenamefont {Dayi}\ \emph {et~al.}(2017)\citenamefont {Dayi},
  \citenamefont {Kilincarslan},\ and\ \citenamefont {Yunt}}]{Dayi:2016foz}%
  \BibitemOpen
  \bibfield  {author} {\bibinfo {author} {\bibfnamefont {O.~F.}\ \bibnamefont
  {Dayi}}, \bibinfo {author} {\bibfnamefont {E.}~\bibnamefont {Kilincarslan}},
  \ and\ \bibinfo {author} {\bibfnamefont {E.}~\bibnamefont {Yunt}},\ }\href
  {\doibase 10.1103/PhysRevD.95.085005} {\bibfield  {journal} {\bibinfo
  {journal} {Phys. Rev.}\ }\textbf {\bibinfo {volume} {D95}},\ \bibinfo {pages}
  {085005} (\bibinfo {year} {2017})},\ \Eprint
  {http://arxiv.org/abs/1605.05451} {arXiv:1605.05451 [hep-th]} \BibitemShut
  {NoStop}%
\bibitem [{\citenamefont {Huang}\ and\ \citenamefont
  {Sadofyev}(2019)}]{Huang:2018aly}%
  \BibitemOpen
  \bibfield  {author} {\bibinfo {author} {\bibfnamefont {X.-G.}\ \bibnamefont
  {Huang}}\ and\ \bibinfo {author} {\bibfnamefont {A.~V.}\ \bibnamefont
  {Sadofyev}},\ }\href {\doibase 10.1007/JHEP03(2019)084} {\bibfield  {journal}
  {\bibinfo  {journal} {JHEP}\ }\textbf {\bibinfo {volume} {03}},\ \bibinfo
  {pages} {084} (\bibinfo {year} {2019})},\ \Eprint
  {http://arxiv.org/abs/1805.08779} {arXiv:1805.08779 [hep-th]} \BibitemShut
  {NoStop}%
\bibitem [{\citenamefont {Winter}(1985)}]{Winter:1986da}%
  \BibitemOpen
  \bibfield  {author} {\bibinfo {author} {\bibfnamefont {J.}~\bibnamefont
  {Winter}},\ }\href {\doibase 10.1103/PhysRevD.32.1871} {\bibfield  {journal}
  {\bibinfo  {journal} {Phys. Rev.}\ }\textbf {\bibinfo {volume} {D32}},\
  \bibinfo {pages} {1871} (\bibinfo {year} {1985})}\BibitemShut {NoStop}%
\bibitem [{\citenamefont {Calzetta}\ \emph {et~al.}(1988)\citenamefont
  {Calzetta}, \citenamefont {Habib},\ and\ \citenamefont
  {Hu}}]{Calzetta:1987bw}%
  \BibitemOpen
  \bibfield  {author} {\bibinfo {author} {\bibfnamefont {E.}~\bibnamefont
  {Calzetta}}, \bibinfo {author} {\bibfnamefont {S.}~\bibnamefont {Habib}}, \
  and\ \bibinfo {author} {\bibfnamefont {B.~L.}\ \bibnamefont {Hu}},\ }\href
  {\doibase 10.1103/PhysRevD.37.2901} {\bibfield  {journal} {\bibinfo
  {journal} {Phys. Rev.}\ }\textbf {\bibinfo {volume} {D37}},\ \bibinfo {pages}
  {2901} (\bibinfo {year} {1988})}\BibitemShut {NoStop}%
\bibitem [{\citenamefont {Fonarev}(1994)}]{Fonarev:1993ht}%
  \BibitemOpen
  \bibfield  {author} {\bibinfo {author} {\bibfnamefont {O.~A.}\ \bibnamefont
  {Fonarev}},\ }\href {\doibase 10.1063/1.530542} {\bibfield  {journal}
  {\bibinfo  {journal} {J. Math. Phys.}\ }\textbf {\bibinfo {volume} {35}},\
  \bibinfo {pages} {2105} (\bibinfo {year} {1994})},\ \Eprint
  {http://arxiv.org/abs/gr-qc/9309005} {arXiv:gr-qc/9309005 [gr-qc]}
  \BibitemShut {NoStop}%
\bibitem [{\citenamefont {Elze}\ \emph {et~al.}(1986)\citenamefont {Elze},
  \citenamefont {Gyulassy},\ and\ \citenamefont {Vasak}}]{Elze:1986qd}%
  \BibitemOpen
  \bibfield  {author} {\bibinfo {author} {\bibfnamefont {H.~T.}\ \bibnamefont
  {Elze}}, \bibinfo {author} {\bibfnamefont {M.}~\bibnamefont {Gyulassy}}, \
  and\ \bibinfo {author} {\bibfnamefont {D.}~\bibnamefont {Vasak}},\ }\href
  {\doibase 10.1016/0550-3213(86)90072-6} {\bibfield  {journal} {\bibinfo
  {journal} {Nucl. Phys.}\ }\textbf {\bibinfo {volume} {B276}},\ \bibinfo
  {pages} {706} (\bibinfo {year} {1986})}\BibitemShut {NoStop}%
\bibitem [{\citenamefont {Heinz}(1983)}]{Heinz:1983nx}%
  \BibitemOpen
  \bibfield  {author} {\bibinfo {author} {\bibfnamefont {U.~W.}\ \bibnamefont
  {Heinz}},\ }\href {\doibase 10.1103/PhysRevLett.51.351} {\bibfield  {journal}
  {\bibinfo  {journal} {Phys. Rev. Lett.}\ }\textbf {\bibinfo {volume} {51}},\
  \bibinfo {pages} {351} (\bibinfo {year} {1983})}\BibitemShut {NoStop}%
\bibitem [{\citenamefont {Nakahara}(2003)}]{nakahara2003geometry}%
  \BibitemOpen
  \bibfield  {author} {\bibinfo {author} {\bibfnamefont {M.}~\bibnamefont
  {Nakahara}},\ }\href@noop {} {\emph {\bibinfo {title} {Geometry, topology and
  physics}}}\ (\bibinfo  {publisher} {CRC Press},\ \bibinfo {year}
  {2003})\BibitemShut {NoStop}%
\bibitem [{\citenamefont {Vasak}\ \emph {et~al.}(1987)\citenamefont {Vasak},
  \citenamefont {Gyulassy},\ and\ \citenamefont {Elze}}]{Vasak:1987um}%
  \BibitemOpen
  \bibfield  {author} {\bibinfo {author} {\bibfnamefont {D.}~\bibnamefont
  {Vasak}}, \bibinfo {author} {\bibfnamefont {M.}~\bibnamefont {Gyulassy}}, \
  and\ \bibinfo {author} {\bibfnamefont {H.~T.}\ \bibnamefont {Elze}},\ }\href
  {\doibase 10.1016/0003-4916(87)90169-2} {\bibfield  {journal} {\bibinfo
  {journal} {Annals Phys.}\ }\textbf {\bibinfo {volume} {173}},\ \bibinfo
  {pages} {462} (\bibinfo {year} {1987})}\BibitemShut {NoStop}%
\bibitem [{\citenamefont {Mathisson}(1937)}]{Mathisson:1937zz}%
  \BibitemOpen
  \bibfield  {author} {\bibinfo {author} {\bibfnamefont {M.}~\bibnamefont
  {Mathisson}},\ }\href@noop {} {\bibfield  {journal} {\bibinfo  {journal}
  {Acta Phys. Polon.}\ }\textbf {\bibinfo {volume} {6}},\ \bibinfo {pages}
  {163} (\bibinfo {year} {1937})}\BibitemShut {NoStop}%
\bibitem [{\citenamefont {Mathisson}(2010)}]{mathisson2010republication}%
  \BibitemOpen
  \bibfield  {author} {\bibinfo {author} {\bibfnamefont {M.}~\bibnamefont
  {Mathisson}},\ }\href@noop {} {\bibfield  {journal} {\bibinfo  {journal}
  {General relativity and gravitation}\ }\textbf {\bibinfo {volume} {42}},\
  \bibinfo {pages} {1011} (\bibinfo {year} {2010})}\BibitemShut {NoStop}%
\bibitem [{\citenamefont {Papapetrou}(1951)}]{Papapetrou:1951pa}%
  \BibitemOpen
  \bibfield  {author} {\bibinfo {author} {\bibfnamefont {A.}~\bibnamefont
  {Papapetrou}},\ }\href {\doibase 10.1098/rspa.1951.0200} {\bibfield
  {journal} {\bibinfo  {journal} {Proc. Roy. Soc. Lond.}\ }\textbf {\bibinfo
  {volume} {A209}},\ \bibinfo {pages} {248} (\bibinfo {year}
  {1951})}\BibitemShut {NoStop}%
\bibitem [{\citenamefont {Parker}\ and\ \citenamefont
  {Toms}(2009)}]{parker2009quantum}%
  \BibitemOpen
  \bibfield  {author} {\bibinfo {author} {\bibfnamefont {L.}~\bibnamefont
  {Parker}}\ and\ \bibinfo {author} {\bibfnamefont {D.}~\bibnamefont {Toms}},\
  }\href@noop {} {\emph {\bibinfo {title} {Quantum field theory in curved
  spacetime: quantized fields and gravity}}}\ (\bibinfo  {publisher} {Cambridge
  University Press},\ \bibinfo {year} {2009})\BibitemShut {NoStop}%
\bibitem [{\citenamefont {Flachi}\ and\ \citenamefont
  {Fukushima}(2018)}]{Flachi:2017vlp}%
  \BibitemOpen
  \bibfield  {author} {\bibinfo {author} {\bibfnamefont {A.}~\bibnamefont
  {Flachi}}\ and\ \bibinfo {author} {\bibfnamefont {K.}~\bibnamefont
  {Fukushima}},\ }\href {\doibase 10.1103/PhysRevD.98.096011} {\bibfield
  {journal} {\bibinfo  {journal} {Phys. Rev.}\ }\textbf {\bibinfo {volume}
  {D98}},\ \bibinfo {pages} {096011} (\bibinfo {year} {2018})},\ \Eprint
  {http://arxiv.org/abs/1702.04753} {arXiv:1702.04753 [hep-th]} \BibitemShut
  {NoStop}%
\bibitem [{\citenamefont {Ebihara}\ \emph
  {et~al.}(2017{\natexlab{a}})\citenamefont {Ebihara}, \citenamefont
  {Fukushima},\ and\ \citenamefont {Mameda}}]{Ebihara:2016fwa}%
  \BibitemOpen
  \bibfield  {author} {\bibinfo {author} {\bibfnamefont {S.}~\bibnamefont
  {Ebihara}}, \bibinfo {author} {\bibfnamefont {K.}~\bibnamefont {Fukushima}},
  \ and\ \bibinfo {author} {\bibfnamefont {K.}~\bibnamefont {Mameda}},\ }\href
  {\doibase 10.1016/j.physletb.2016.11.010} {\bibfield  {journal} {\bibinfo
  {journal} {Phys. Lett.}\ }\textbf {\bibinfo {volume} {B764}},\ \bibinfo
  {pages} {94} (\bibinfo {year} {2017}{\natexlab{a}})},\ \Eprint
  {http://arxiv.org/abs/1608.00336} {arXiv:1608.00336 [hep-ph]} \BibitemShut
  {NoStop}%
\bibitem [{\citenamefont {Skagerstam}(1992)}]{Skagerstam:1992er}%
  \BibitemOpen
  \bibfield  {author} {\bibinfo {author} {\bibfnamefont {B.~S.}\ \bibnamefont
  {Skagerstam}},\ }\href@noop {} {\  (\bibinfo {year} {1992})},\ \Eprint
  {http://arxiv.org/abs/hep-th/9210054} {arXiv:hep-th/9210054 [hep-th]}
  \BibitemShut {NoStop}%
\bibitem [{\citenamefont {Stone}\ \emph {et~al.}(2015)\citenamefont {Stone},
  \citenamefont {Dwivedi},\ and\ \citenamefont {Zhou}}]{Stone:2015kla}%
  \BibitemOpen
  \bibfield  {author} {\bibinfo {author} {\bibfnamefont {M.}~\bibnamefont
  {Stone}}, \bibinfo {author} {\bibfnamefont {V.}~\bibnamefont {Dwivedi}}, \
  and\ \bibinfo {author} {\bibfnamefont {T.}~\bibnamefont {Zhou}},\ }\href
  {\doibase 10.1103/PhysRevLett.114.210402} {\bibfield  {journal} {\bibinfo
  {journal} {Phys. Rev. Lett.}\ }\textbf {\bibinfo {volume} {114}},\ \bibinfo
  {pages} {210402} (\bibinfo {year} {2015})},\ \Eprint
  {http://arxiv.org/abs/1501.04586} {arXiv:1501.04586 [hep-th]} \BibitemShut
  {NoStop}%
\bibitem [{\citenamefont {Dayi}\ and\ \citenamefont
  {Kilincarslan}(2018)}]{Dayi:2018xdy}%
  \BibitemOpen
  \bibfield  {author} {\bibinfo {author} {\bibfnamefont {O.~F.}\ \bibnamefont
  {Dayi}}\ and\ \bibinfo {author} {\bibfnamefont {E.}~\bibnamefont
  {Kilincarslan}},\ }\href {\doibase 10.1103/PhysRevD.98.081701} {\bibfield
  {journal} {\bibinfo  {journal} {Phys. Rev.}\ }\textbf {\bibinfo {volume}
  {D98}},\ \bibinfo {pages} {081701} (\bibinfo {year} {2018})},\ \Eprint
  {http://arxiv.org/abs/1807.05912} {arXiv:1807.05912 [hep-th]} \BibitemShut
  {NoStop}%
\bibitem [{\citenamefont {Gao}\ \emph {et~al.}(2018)\citenamefont {Gao},
  \citenamefont {Pang},\ and\ \citenamefont {Wang}}]{Gao:2018jsi}%
  \BibitemOpen
  \bibfield  {author} {\bibinfo {author} {\bibfnamefont {J.-H.}\ \bibnamefont
  {Gao}}, \bibinfo {author} {\bibfnamefont {J.-Y.}\ \bibnamefont {Pang}}, \
  and\ \bibinfo {author} {\bibfnamefont {Q.}~\bibnamefont {Wang}},\ }\href@noop
  {} {\  (\bibinfo {year} {2018})},\ \Eprint {http://arxiv.org/abs/1810.02028}
  {arXiv:1810.02028 [nucl-th]} \BibitemShut {NoStop}%
\bibitem [{\citenamefont {Janka}(2012)}]{Janka:2012wk}%
  \BibitemOpen
  \bibfield  {author} {\bibinfo {author} {\bibfnamefont {H.-T.}\ \bibnamefont
  {Janka}},\ }\href {\doibase 10.1146/annurev-nucl-102711-094901} {\bibfield
  {journal} {\bibinfo  {journal} {Ann. Rev. Nucl. Part. Sci.}\ }\textbf
  {\bibinfo {volume} {62}},\ \bibinfo {pages} {407} (\bibinfo {year} {2012})},\
  \Eprint {http://arxiv.org/abs/1206.2503} {arXiv:1206.2503 [astro-ph.SR]}
  \BibitemShut {NoStop}%
\bibitem [{\citenamefont {Yamamoto}(2016)}]{Yamamoto:2015gzz}%
  \BibitemOpen
  \bibfield  {author} {\bibinfo {author} {\bibfnamefont {N.}~\bibnamefont
  {Yamamoto}},\ }\href {\doibase 10.1103/PhysRevD.93.065017} {\bibfield
  {journal} {\bibinfo  {journal} {Phys. Rev.}\ }\textbf {\bibinfo {volume}
  {D93}},\ \bibinfo {pages} {065017} (\bibinfo {year} {2016})},\ \Eprint
  {http://arxiv.org/abs/1511.00933} {arXiv:1511.00933 [astro-ph.HE]}
  \BibitemShut {NoStop}%
\bibitem [{\citenamefont {Jiang}\ \emph {et~al.}(2016)\citenamefont {Jiang},
  \citenamefont {Lin},\ and\ \citenamefont {Liao}}]{Jiang:2016woz}%
  \BibitemOpen
  \bibfield  {author} {\bibinfo {author} {\bibfnamefont {Y.}~\bibnamefont
  {Jiang}}, \bibinfo {author} {\bibfnamefont {Z.-W.}\ \bibnamefont {Lin}}, \
  and\ \bibinfo {author} {\bibfnamefont {J.}~\bibnamefont {Liao}},\ }\href
  {\doibase 10.1103/PhysRevC.94.044910, 10.1103/PhysRevC.95.049904} {\bibfield
  {journal} {\bibinfo  {journal} {Phys. Rev.}\ }\textbf {\bibinfo {volume}
  {C94}},\ \bibinfo {pages} {044910} (\bibinfo {year} {2016})},\ \bibinfo
  {note} {[Erratum: Phys. Rev.C95,no.4,049904(2017)]},\ \Eprint
  {http://arxiv.org/abs/1602.06580} {arXiv:1602.06580 [hep-ph]} \BibitemShut
  {NoStop}%
\bibitem [{\citenamefont {Deng}\ and\ \citenamefont
  {Huang}(2016)}]{Deng:2016gyh}%
  \BibitemOpen
  \bibfield  {author} {\bibinfo {author} {\bibfnamefont {W.-T.}\ \bibnamefont
  {Deng}}\ and\ \bibinfo {author} {\bibfnamefont {X.-G.}\ \bibnamefont
  {Huang}},\ }\href {\doibase 10.1103/PhysRevC.93.064907} {\bibfield  {journal}
  {\bibinfo  {journal} {Phys. Rev.}\ }\textbf {\bibinfo {volume} {C93}},\
  \bibinfo {pages} {064907} (\bibinfo {year} {2016})},\ \Eprint
  {http://arxiv.org/abs/1603.06117} {arXiv:1603.06117 [nucl-th]} \BibitemShut
  {NoStop}%
\bibitem [{\citenamefont {Ebihara}\ \emph
  {et~al.}(2017{\natexlab{b}})\citenamefont {Ebihara}, \citenamefont
  {Fukushima},\ and\ \citenamefont {Pu}}]{Ebihara:2017suq}%
  \BibitemOpen
  \bibfield  {author} {\bibinfo {author} {\bibfnamefont {S.}~\bibnamefont
  {Ebihara}}, \bibinfo {author} {\bibfnamefont {K.}~\bibnamefont {Fukushima}},
  \ and\ \bibinfo {author} {\bibfnamefont {S.}~\bibnamefont {Pu}},\ }\href
  {\doibase 10.1103/PhysRevD.96.016016} {\bibfield  {journal} {\bibinfo
  {journal} {Phys. Rev.}\ }\textbf {\bibinfo {volume} {D96}},\ \bibinfo {pages}
  {016016} (\bibinfo {year} {2017}{\natexlab{b}})},\ \Eprint
  {http://arxiv.org/abs/1705.08611} {arXiv:1705.08611 [hep-ph]} \BibitemShut
  {NoStop}%
\bibitem [{\citenamefont {Luttinger}(1964)}]{Luttinger:1964zz}%
  \BibitemOpen
  \bibfield  {author} {\bibinfo {author} {\bibfnamefont {J.~M.}\ \bibnamefont
  {Luttinger}},\ }\href {\doibase 10.1103/PhysRev.135.A1505} {\bibfield
  {journal} {\bibinfo  {journal} {Phys. Rev.}\ }\textbf {\bibinfo {volume}
  {135}},\ \bibinfo {pages} {A1505} (\bibinfo {year} {1964})}\BibitemShut
  {NoStop}%
\bibitem [{\citenamefont {Hayata}\ \emph {et~al.}()\citenamefont {Hayata},
  \citenamefont {Hidaka},\ and\ \citenamefont {Mameda}}]{Mameda:pre}%
  \BibitemOpen
  \bibfield  {author} {\bibinfo {author} {\bibfnamefont {T.}~\bibnamefont
  {Hayata}}, \bibinfo {author} {\bibfnamefont {Y.}~\bibnamefont {Hidaka}}, \
  and\ \bibinfo {author} {\bibfnamefont {K.}~\bibnamefont {Mameda}},\
  }\href@noop {} {}\bibinfo {note} {{in preparation}}\BibitemShut {NoStop}%
\bibitem [{\citenamefont {Cortijo}\ \emph {et~al.}(2015)\citenamefont
  {Cortijo}, \citenamefont {Ferreiros}, \citenamefont {Landsteiner},\ and\
  \citenamefont {Vozmediano}}]{PhysRevLett.115.177202}%
  \BibitemOpen
  \bibfield  {author} {\bibinfo {author} {\bibfnamefont {A.}~\bibnamefont
  {Cortijo}}, \bibinfo {author} {\bibfnamefont {Y.}~\bibnamefont {Ferreiros}},
  \bibinfo {author} {\bibfnamefont {K.}~\bibnamefont {Landsteiner}}, \ and\
  \bibinfo {author} {\bibfnamefont {M.~A.~H.}\ \bibnamefont {Vozmediano}},\
  }\href {\doibase 10.1103/PhysRevLett.115.177202} {\bibfield  {journal}
  {\bibinfo  {journal} {Phys. Rev. Lett.}\ }\textbf {\bibinfo {volume} {115}},\
  \bibinfo {pages} {177202} (\bibinfo {year} {2015})},\ \Eprint
  {http://arxiv.org/abs/1603.02674} {arXiv:1603.02674 [cond-mat.mes-hall]}
  \BibitemShut {NoStop}%
\bibitem [{\citenamefont {Cortijo}\ \emph {et~al.}(2016)\citenamefont
  {Cortijo}, \citenamefont {Kharzeev}, \citenamefont {Landsteiner},\ and\
  \citenamefont {Vozmediano}}]{Cortijo:2016wnf}%
  \BibitemOpen
  \bibfield  {author} {\bibinfo {author} {\bibfnamefont {A.}~\bibnamefont
  {Cortijo}}, \bibinfo {author} {\bibfnamefont {D.}~\bibnamefont {Kharzeev}},
  \bibinfo {author} {\bibfnamefont {K.}~\bibnamefont {Landsteiner}}, \ and\
  \bibinfo {author} {\bibfnamefont {M.~A.~H.}\ \bibnamefont {Vozmediano}},\
  }\href {\doibase 10.1103/PhysRevB.94.241405} {\bibfield  {journal} {\bibinfo
  {journal} {Phys. Rev.}\ }\textbf {\bibinfo {volume} {B94}},\ \bibinfo {pages}
  {241405} (\bibinfo {year} {2016})},\ \Eprint
  {http://arxiv.org/abs/1607.03491} {arXiv:1607.03491 [cond-mat.mes-hall]}
  \BibitemShut {NoStop}%
\bibitem [{\citenamefont {Grushin}\ \emph {et~al.}(2016)\citenamefont
  {Grushin}, \citenamefont {Venderbos}, \citenamefont {Vishwanath},\ and\
  \citenamefont {Ilan}}]{PhysRevX.6.041046}%
  \BibitemOpen
  \bibfield  {author} {\bibinfo {author} {\bibfnamefont {A.~G.}\ \bibnamefont
  {Grushin}}, \bibinfo {author} {\bibfnamefont {J.~W.~F.}\ \bibnamefont
  {Venderbos}}, \bibinfo {author} {\bibfnamefont {A.}~\bibnamefont
  {Vishwanath}}, \ and\ \bibinfo {author} {\bibfnamefont {R.}~\bibnamefont
  {Ilan}},\ }\href {\doibase 10.1103/PhysRevX.6.041046} {\bibfield  {journal}
  {\bibinfo  {journal} {Phys. Rev. X}\ }\textbf {\bibinfo {volume} {6}},\
  \bibinfo {pages} {041046} (\bibinfo {year} {2016})}\BibitemShut {NoStop}%
\bibitem [{\citenamefont {Sumiyoshi}\ and\ \citenamefont
  {Fujimoto}(2016)}]{Sumiyoshi:2015eda}%
  \BibitemOpen
  \bibfield  {author} {\bibinfo {author} {\bibfnamefont {H.}~\bibnamefont
  {Sumiyoshi}}\ and\ \bibinfo {author} {\bibfnamefont {S.}~\bibnamefont
  {Fujimoto}},\ }\href {\doibase 10.1103/PhysRevLett.116.166601} {\bibfield
  {journal} {\bibinfo  {journal} {Phys. Rev. Lett.}\ }\textbf {\bibinfo
  {volume} {116}},\ \bibinfo {pages} {166601} (\bibinfo {year} {2016})},\
  \Eprint {http://arxiv.org/abs/1509.03981} {arXiv:1509.03981
  [cond-mat.mes-hall]} \BibitemShut {NoStop}%
\bibitem [{\citenamefont {DeWitt}(2011)}]{dewitt2011bryce}%
  \BibitemOpen
  \bibfield  {author} {\bibinfo {author} {\bibfnamefont {B.}~\bibnamefont
  {DeWitt}},\ }\href@noop {} {\emph {\bibinfo {title} {Bryce DeWitt's Lectures
  on Gravitation: Edited by Steven M. Christensen}}}\ (\bibinfo  {publisher}
  {Springer},\ \bibinfo {year} {2011})\BibitemShut {NoStop}%
\end{thebibliography}%

\end{document}